\documentclass[aps,prb,twocolumn,groupedaddress,superscriptaddress]{revtex4}

\usepackage{amssymb}
\usepackage{graphicx}
\usepackage{amsmath,color}
\usepackage{feynmf}

\newcommand{\comment}[1]{}
\newcommand{\gv}[1]{{\underline{#1}}}
\newcommand{\s}{\sigma}

\renewcommand{\o}{\omega}
\newcommand{\e}{\epsilon}
\renewcommand{\k}{\gv{k}}

\begin{document}

%\preprint{}

\title{Non-Fermi liquid signatures in the Hubbard Model due to van Hove singularities}


\author{Sebastian Schmitt}
\email{sebastian.schmitt@tu-dortmund.de}
\affiliation{Lehrstuhl f\"{u}r Theoretische Physik II, Technische Universit\"{a}t Dortmund,
Otto-Hahn Str. 4, D-44221 Dortmund, Germany}

%\affiliation{Institut f\"{u}r Festk\"{o}rperphysik, Technische Universit\"{a}t Darmstadt,
%Hochschulstr. 6, D-64289 Darmstadt, Germany}


\date{\today}

\begin{abstract}
  When  a van-Hove singularity is located in the vicinity of the Fermi level, the
  electronic scattering rate acquires
  a non-analytic contribution.
  This invalidates basic assumptions of Fermi liquid theory and  within 
  treatments based on perturbation theory leads to 
  a non-Fermi liquid 
  self-energy and  transport properties.
  Such anomalies are shown to also occur in the  strongly correlated 
  metallic state within dynamical mean-field theory. We consider the Hubbard model on a two-dimensional 
  square lattice with nearest and next-nearest neighbor hopping within the single-site 
  dynamical mean-field theory. At temperatures on the order of the
  low-energy scale $T_0$ an unusual maximum emerges in the imaginary part of the 
  self-energy which is renormalized towards the Fermi level for finite doping.  
  At zero temperature this double-well structure is suppressed, but an anomalous 
  energy dependence of the self-energy remains. For the frustrated Hubbard model on 
  the square lattice with next-nearest neighbor hopping, the presence of the 
  van Hove singularity changes the asymptotic low temperature behavior
  of the resistivity  from a Fermi liquid to non-Fermi liquid dependency as function of
  doping. The results of this work are discussed regarding
  their relevance for high-temperature cuprate superconductors.
\end{abstract}

\pacs{qs}
%\keywords{}

\maketitle

\section{Introduction}
\label{sec:intro}

The question whether interacting fermions
form a Landau Fermi liquid  state
at low temperatures or rather transform into something
commonly referred to as singular or non-Fermi liquid
is a very fundamental one.
A theoretical description requires at least 
a qualitative understanding of the 
low temperature phase 
in order to address the relevant degrees of freedom in that phase.
But for strongly interacting electronic systems this information is 
usually lacking, and the low-energy degrees of freedom might be 
very different from free electrons. 
However, 
the  Fermi liquid picture is often
appropriate 
at low temperatures and energies.
Reasons for this can be found in the stability of the Fermi liquid fixed point
of the renormalization group analysis\cite{shankarRGReview94}
or phase space arguments.\cite{abrikosovBook63}
In both approaches, the analyticity of the scattering vertices 
is crucial for the arguments. 

In situations, where the Fermi liquid description fails, 
the origin for this is linked to low-dimensionality of the system,
long-range interactions,
nested Fermi surfaces or the proximity to a quantum critical 
point, just to name a few (see, e.g.\ Refs.\ \onlinecite{virosztekNestedFL90,Voit1995,varmaSingularFL02,vojtaQPT03}).
The quasiparticle concept might still be valid, but
the scattering rate is strongly enhanced and 
shows an unusual, i.e.\ non-quadratic, 
temperature and frequency dependence. Alternatively  the quasiparticle concept
may also break down completely.

In this work we will focus on 
the single-band Hubbard model which is a minimal model for
strongly correlated electron systems, and 
obtain a non-perturbative approximative solution 
by means of the dynamical mean-field theory (DMFT).\cite{pruschke:dmftNCA_HM95,georges:dmft96}
Within DMFT, where only  local  correlations are included, 
the  generic low temperature state of the metallic Hubbard model is 
a Fermi liquid.\cite{muellerHartmannDInfty89-2,jarrell:HM93b,pruschke:HMTransport93,pruschke:dmftNCA_HM95,bulla:MITHM99,merinoTransportDMFT00}
This is expected, since the low temperature phase of the spin-$\frac{1}{2}$ 
single impurity Anderson model --- the model onto which the Hubbard model is 
mapped within DMFT --- is a local Fermi liquid under the assumption 
of  a ``well behaved'' medium.\cite{hewson:KondoProblem93}
The Fermi liquid reveals itself via a low-energy many-body bandstructure
forming  around the Fermi level
at low temperatures.\cite{grewe:quasipart05,greweCA108}
Due to the neglect of momentum dependent correlations in DMFT 
the heavy quasiparticles possess the non-interacting Fermi 
surface.\cite{muellerHartmannDInfty89-2} 
A characteristic low-energy scale $T_0$  is associated with 
this lattice version of the Kondo effect
and marks the temperature scale for local-moment screening and the emergence 
of coherent quasiparticles.

On the other hand it is known from perturbation theory and
renormalization group treatments,
that a van Hove singularity of the non-interacting density of states (DOS)
in the vicinity of the Fermi level lead to a marginal\cite{varmaMarginaFL89} or non-Fermi liquid 
form of 
the self-energy.\cite{newnsVanHove91,newnsRev92,gopalanSaddle92,hlubninaHotSpot95,dzyaloshinskiiVanHove96,markiewiczHoveRev97,
menashe2DHove99,onufrievaVanHove99,
kastrinakisVanHove00,irkhinFL01,katanin:RGMagSCInstabl03,kataninQPRG04,roheNFL05,rolandFS06}
This should be contrasted to  nested Fermi liquids,\cite{virosztekNestedFL90} where
due to the nesting property of the Fermi surface 
the phase space volume for low-energy   scattering
is also strongly enhanced and unusual low-energy properties emerge as well.\cite{virosztekNestedFL90,schlottmannNFLNesting03} 

Van Hove singularities are a consequence of maxima or
saddle points in the dispersion  relation $t_\k$.
For periodic energy bands in a crystal a certain number of these 
van Hove singularities must occur on topological grounds (see, 
e.g.\ Ref. \onlinecite{jonesMarchBook73}).
For example, all of the cubic lattices have noticeable van Hove singularities in 
space dimensions $d\lesssim 3$,  typically logarithmic divergences 
or square-root cusps.

Anomalous low-energy behavior can already be anticipated by 
recognizing, that the usual arguments of 
microscopic Fermi liquid theory break down if a maximum or saddle point 
of the dispersion relation $t_\k$ is found at the Fermi level.
Then, $t_\k$ \textit{cannot} be approximated by a linear
dispersion, i.e.\
$t_\k\not\approx \frac{k_F}{m^*}|\Delta\k|$,
with $\Delta\k=\k-\k_F$ and  $\k_F$ the Fermi wave vector.
Around these points
a quadratic (maximum) or hyperbolic (saddle-point) 
functional dependence 
$  t_\k\sim   \Delta k_\perp^2\pm \Delta k_\parallel^2$
results, where $k_\perp$ and $k_\parallel$ are two
linear independent 
directions in momentum space.
The arguments leading to a quadratic Fermi liquid
energy and temperature dependence of the scattering rate, i.e.\ 
Im$\Sigma^U(\o-i0^+)\sim \o^2+\pi^2T^2$,
are thus not applicable.
Instead, the phase space volume for scattering is strongly 
enhanced and anomalous $T$- and $\omega$-dependencies may 
result.
If the singularity occurs  in the vicinity of the Fermi level,
strong modifications of the low-energy and temperature 
properties are still expected due to a
pinning effect.\cite{siLinRes91,krishnamurthyHove2d96,gonzalesVanHove97,kastrinakisVanHove00,irkhinVanHove02,odashimaPinn05}

In the present work we show  that  the 
non-Fermi liquid signatures which were encountered in previous 
studies\cite{newnsVanHove91,newnsRev92,gopalanSaddle92,hlubninaHotSpot95,dzyaloshinskiiVanHove96,markiewiczHoveRev97,
menashe2DHove99,onufrievaVanHove99,
kastrinakisVanHove00,irkhinFL01,katanin:RGMagSCInstabl03,kataninQPRG04,roheNFL05,rolandFS06}
of the self-energy  and transport properties
can also occur in the paramagnetic phase of a strongly correlated metal within DMFT.
We use the two-dimensional  square lattice with nearest and next-nearest 
neighbor hopping, $t$ and  $t'$ respectively.
For $-0.5< t'/t< 0.5$ the dispersion relation $t_\k$ has saddles at 
the $X$ points in the Brillouin zone 
($\gv{k}_X=\{(0,\pm\pi),(\pm\pi,0)\}$). 
This leads to a logarithmic divergence in the non-interacting DOS, which
is located at the Fermi level $\o=0$ for $t'=0$ and moved to finite energies
for $t'\neq 0$.

The resulting anomalies consist of an unusual maximum in the imaginary part of the 
self-energy Im$\Sigma^U(\o-i0^+)$ in the vicinity of the Fermi level 
at low but finite temperatures. 
The concept of  quasiparticles remains valid but they 
are  strongly scattered at the 
flat parts of the dispersion relation and   the Fermi liquid formation
is disturbed.
The anomalous  double well structure persists
over a range of doping and next-nearest neighbor hopping $t'$.   
At zero temperature and for half-filled  square lattice 
the double-well structure is removed
and the self-energy vanishes at the Fermi level.
Even though the self-energy exhibits a non-Fermi liquid energy
dependence, the quasiparticle weight is still finite.
The  pinning of the van Hove singularity to the Fermi level, which has previously 
been found,\cite{siLinRes91,newnsVanHove91,krishnamurthyHove2d96,gonzalesVanHove97,kastrinakisVanHove00,irkhinVanHove02,odashimaPinn05}
is explicitly observed in the correlated metal.

Qualitatively similar non-Fermi liquid signatures  are also observed in more advanced 
theories which include nonlocal correlations.\cite{maier:QuantumCluster05,macridinCupratesTwobandHM05,
   tremblayPseudogap06,macridinPseudogapHM06,kyungPseudogap06,liebsch2DHM09,kataninDGA2dHM09}
These
features are usually attributed 
to the presence of strong nonlocal antiferromagnetic correlations.
However, in the present approach 
nonlocal  correlations are not included and we argue that
an interpretation \textit{exclusively} 
in terms of  nonlocal  correlations is too simplified.

The paper is organized as follows. 
In Section~\ref{sec:theo} we shortly present the Hubbard Hamiltonian
and the basic equations of DMFT.   
We elucidate the role of the non-analyticities in the non-interacting DOS
by means of a simplified model DOS for which analytical insights are obtained
in Sect.~\ref{sec:hubNFLcusp}. 
Self-consistent DMFT calculations for singular and non-singular model DOS reveal 
the non-analyticity to be the cause for the observed non-Fermi liquid signatures. 
In Sect.~\ref{sec:square} we present finite and zero temperature calculations for the spectral function
and self-energy for a square lattice without
next-nearest neighbor hopping $t'=0$. The anomalies are shown to occur at temperatures on the
order of the low  energy scale $T_0$ and decrease with temperature. At $T=0$ a 
non-quadratic energy dependence in the imaginary part of the self-energy remains.
The case of the square lattice with next-nearest neighbor hopping $t'=-0.2t$
is discussed in Sect.~\ref{sec:ttp}. In addition to the 
spectral function and the self-energy, we also study the temperature dependence of the
zero frequency quasiparticle scattering rate and the resistivity. 
We discuss in Sect.~\ref{sec:susNFLRelPG} 
the relevance of the present findings to the cuprate superconductors
before we conclude
the paper.

%%%%%%%%%%%%%%%%%%%%%%%%%%%%%%%%%%%
\section{Model and Method}
\label{sec:theo}

We study the single band  Hubbard model
\begin{align}
  \label{eq:ham}
  \hat H&=
  \sum_{ij,\s}t_{ij}\hat{c}^\dagger_{i\s}\hat{c}_{j\s}
  +\sum_{i\,\s} \e \:\hat{c}^\dagger_{i\s}\hat{c}_{i\s}
  +U\, \sum_i \hat{n}_{i\uparrow}\hat{n}_{i\downarrow}
  \quad,
\end{align}
where the operator $\hat{c}_{i\s}$ ($\hat{c}^\dagger_{i\s}$) 
annihilates (creates) an electron in a 
localized Wannier orbital at lattice  site $i$ with spin $\s$, 
$\hat{n}_{i\s}=\hat{c}^\dagger_{i\s}\hat{c}_{i\s}$ 
represents the number operator for electrons, $\e$ is the 
local ionic level position at each site,
and $U$ is the local matrix element of the Coulomb repulsion. 

In the following an external magnetic field  
is not included and all spin up and down
quantities are equal.
The spin label of Green's functions and self-energies will be suppressed.
We focus on
the paramagnetic phase of the model and exclude  possible
phase transitions to ordered phases such 
as superconductivity or magnetism.

All information on the lattice is encoded in the
one-particle hopping amplitude $t_{ij}$. 
Its Fourier-transform gives the dispersion relation
\begin{align}
  \label{eq:disp}
  t_\k&=-2t (\cos k_x+\cos k_y)-4t' \cos k_x\cos k_y
\end{align}
where we set the lattice constant $a=1$ and already specified to the two-dimensional ($2d$)
square lattice with nearest and next-nearest neighbor 
hopping, $t$ and $t'$, respectively.

The central quantity for the discussion of
Fermi liquid properties 
is the correlation self-energy 
$\Sigma^{U}(z)$. The assumption that it does  not depend 
on momentum $\k$ constitutes the major approximation of
DMFT.
The lattice Green's function of the Hubbard model  is then given by
\begin{align}
  \label{eq:gfk}
  G(t_\k,z)=\frac{1}{z-t_\k-\Sigma^U(z)}
  \quad,
\end{align}
which only depends on $\k$ via the dispersion relation $t_\k$.
The latter also determines 
the non-interacting DOS  
$\rho_0(\o)=\frac{1}{N}\sum_\k\delta(\o-t_\k)$.

Within DMFT the lattice model is mapped onto an
effective impurity model, where the medium $\Gamma(z)$ 
of the  impurity model cannot be chosen
arbitrarily, but has to be determined from the self-consistency conditions
\begin{align}
  \label{eq:mediumSC} \Gamma(z)&= \frac{T(z)}{1+\tilde{G}(z)T(z)} \\
  \label{eq:scatter} T(z)& = \frac{1}{N}\sum_\k \frac{ t_\k^2}{\tilde{G}(z)^{-1}-t_\k}
  =  \int\!dx\: \frac{ x^2\:\rho_{0}(x)}{\tilde{G}(z)^{-1}-x}\\
  \label{eq:gtilde}\tilde{G}(z)&= \frac{1}{z-\e-\Sigma^U(z)} 
  \quad.
\end{align}
The functional form of the local scattering matrix $T(z)$ is completely
determined by the 
non-interacting DOS. The correlation self-energy enters only as an
unspecified parameter [via $\tilde G(z)$] and is obtained from the 
solution of the effective impurity model for a given medium $\Gamma(z)$.
The self-consistency cycle is closed by identifying the
local Green's function of the lattice with the one of the impurity model,
\begin{align}
  \label{eq:GlocLatt} G^{\text{(loc)}}(z)&=\frac{1}{z-\e-\Sigma^U(z)-\Gamma(z)} 
  \\  \notag 
  &\overset{!}{=}\frac{1}{N}\sum_\k G(t_\k,z) =\tilde{G}(z)+\tilde{G}(z)\: T(z)\: \tilde{G}(z) 
\quad .
\end{align}

For the finite temperature calculations we use the enhanced non-crossing
approximation\cite{pruschkeENCA89,holmFiniteUNCA89,keiterENCA90,grewe:quasipart05,greweCA108} (ENCA) 
as the impurity solver. There are no adjustable parameters 
and the method works directly on the real frequency axis. 
This is crucial for the observation 
of the small low-energy anomalies in the self-energy. However, it cannot be employed
at too low temperatures $T\ll  T_0$ due to the NCA pathology.\cite{bickers:nca87a}
Within the ENCA the pathologic behavior is considerably improved compared to 
NCA.\cite{schmittSus09} It can be reliably used    down to 
temperatures $T\approx T_0/10$  for moderate to large values of $U$ and not too far away 
from half-filling.  

At zero temperature we use the numerical renormalization 
group\cite{krishnamurtyNRGSIAMI80,krishnamurtyNRGSIAMII80,bullaNRGReview08} (NRG) instead.   
For the NRG spectral functions we use the complete Fock-space\cite{andersRealTimeDynNRG05,andersRealTimeKondo06}
algorithm of Ref.~\onlinecite{petersNewNRG06} with a discretization parameter
$\Lambda=1.7$ and keep approximately  $1700$  states in each NRG-iteration step. 
We average over eight different discretizations\cite{campoZtrickNRG05} of the conduction
band to minimize the errors. 

We also performed finite temperature DMFT(NRG) 
calculations which  were in accord  with the DMFT(ENCA) results 
and confirmed the existence of the anomalous structures in the self-energy.
However, the anomalous features occur at energies and temperatures 
where  the results depend on the actual values of 
NRG broadening parameters.\cite{bulla:MITFTNRG01,petersNewNRG06}
Therefore, we prefer the ENCA for 
the finite temperature calculations.

%%%%%%%%%%%%%%%%%%%%%%%%%%%%%%%%%%%%%%%%%%%%
\section{Model density of states}
\label{sec:hubNFLcusp}

\begin{figure}[t]
  \centering
  \includegraphics[width=80mm]{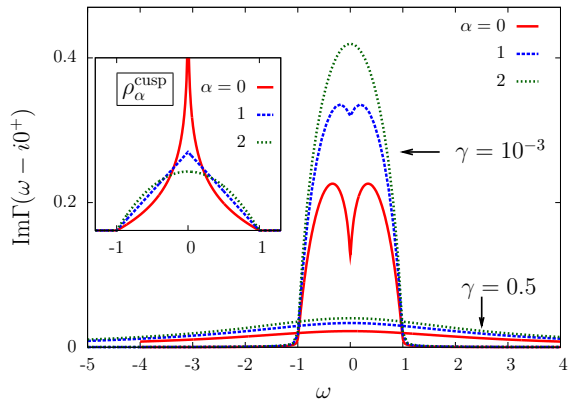}
  \caption{Imaginary parts of the  effective 
    media $\Gamma(\o-i0^+)$ [see Eqs.~\eqref{eq:hubToyLinDos0}, 
    ~\eqref{eq:hubToyLinDos1} and \eqref{eq:hubToyLinDos2}]
    as functions of  energy $\o$  for the 
    cases $\alpha=0,1,2$  and two values of the residual interaction 
    $\gamma$. The inset shows the model DOS
    for the same values of $\alpha$.
    $W=Z=1$ and $T=\tilde\e=0$ was used.
  }
  \label{fig:hubToyDosTtilde}
\end{figure}

Before presenting DMFT results for the Hubbard model on
cubic lattices we elucidate the basic mechanism by means of 
a simplified model DOS of the form
\begin{align}
  \label{eq:hubToyDosCusp}
  \rho^{\text{cusp}}_\alpha(\o)&=
    \frac{1+\alpha}{2\alpha W}\left[1-\left(\frac{|\o|}{W}\right)^\alpha\right] ,\quad  \alpha\geq 0
\end{align}
with $-W\leq\o\leq W$. For $\alpha< 2$  this DOS 
has a non-analytic cusp at $\o=0$ which
turns into a logarithmic  divergence for $\alpha= 0$,
$\rho^{\text{cusp}}_{\alpha=0}(\o)=\frac{1}{2W}\ln\Big(\frac{W}{|\o|}\Big)$,
while it is smooth for $\alpha\geq 2$ (see inset of Fig.~\ref{fig:hubToyDosTtilde}).
The case $\alpha=0$ mimics the logarithmic divergence of the $2d$ simple
cubic lattice.

The explicit form (\ref{eq:hubToyDosCusp}) allows for the  analytic calculation
of the effective medium $\Gamma(z)$ 
as shown in the Appendix.
We state the result for three values of $\alpha$
\begin{align}
  \label{eq:hubToyLinDos0}
  \Gamma_{\alpha=0}(z)&=
  \frac{1}{\tilde G(z)} + \frac{2 W}{\text{Li}_2[-\tilde{G}(z)W] - \text{Li}_2[\tilde{G}(z)W]}
\end{align}
\begin{align}
  \label{eq:hubToyLinDos1}
  \Gamma_{\alpha=1}(z)&=
  \frac{1}{\tilde{G}(z)}
  \\ \notag
  &
  \hfill
  - \frac{W}{
    \frac{1}{\tilde{G}(z) W}\ln\left[1-(\tilde{G}(z)W)^2\right]+2\mathrm{atanh}[\tilde{G}(z) W]
  }
\end{align}
\begin{align}
  \label{eq:hubToyLinDos2}
  \Gamma_{\alpha=2}(z)&=
  \frac{1}{\tilde{G}(z)}
  \\ \notag
  &
  \hfill -
  \frac{(2/3)W}{
    \frac{1}{\tilde{G}(z)W}+\left(1-\frac{1}{(\tilde{G}(z)W)^2}\right)\mathrm{atanh}[\tilde{G}(z) W]
  }    
  \quad,
\end{align}
where Li$_2(z)$ is the Dilogarithm. 
The crucial difference in the analytic structure of the three results is the  appearance of
non-analytic logarithms for $\alpha=0$ and $\alpha=1$  which are absent 
for $\alpha=2$.

Suppose the system is about to form a Fermi liquid state as it is expected 
within DMFT for the Hubbard model at low $T$.
Then, the 
correlation self-energy has the form, 
\begin{align}
\label{eq:flSig}
\o-\epsilon - \Sigma^U(\o-i0^+)=\frac{1}{Z}\left(\o-\tilde \e -i\gamma \left[(\pi T)^2+\o^2\right]\right)
,
\end{align}
where $T$ is the temperature,  $\gamma$ a measure for the residual 
quasiparticle scattering, and $Z$ the quasiparticle weight.
The renormalized single-particle energy $\tilde \e=\e+\text{Re}\Sigma^U(-i0^+)$ 
determines the Fermi wave vector via $\tilde \e=t_{\k_F}$ and vanishes  
in symmetric situations.

Using the  form (\ref{eq:flSig}) in the expressions
Eqs.~(\ref{eq:hubToyLinDos0}) to  (\ref{eq:hubToyLinDos2}),
the imaginary part of $\Gamma$ develops a cusp-like minimum at the 
Fermi level 
for small $\gamma$ and $\alpha<2$.\cite{schmittPhD08}
Figure~\ref{fig:hubToyDosTtilde} illustrates this for the three different 
values of the exponent $\alpha=0,1$, and $2$.
%%%%%%%%%%%%%%%%%%%%%%%%%%%%%%%%%%%%%% 
In the limit $\gamma\to 0$ the effective media for $\alpha=1$ and $\alpha=2$ approach the values
Im$\Gamma_{\alpha=1}=1/\pi$ and Im$\Gamma_{\alpha=2}=4/(3\pi)$, respectively, 
while for $\alpha=0$ the effective medium 
eventually reaches zero (see Eqs.~(\ref{eq:tta0w0}) to~(\ref{eq:tta2w0})).
Thus, for $\alpha=0$, we end up with a soft-gap
effective medium for which Kondo screening  and the local Fermi 
liquid behavior can be completely  destroyed under 
certain conditions.\cite{bullaPseudoGapSIAM97,gonzalesSIAMPseudogap98,loganPseudogapSIAM00,glossopPseudogap00,ingersentPseudogapSIAM02,glossopSIAMPseudogap03,fritzPseudagapSIAM04,fritzPseudogapKondo06}
However, in the present case the effective medium vanishes only logarithmically 
at the Fermi level, i.e.\
Im$\Gamma_{\alpha=0}(\omega-i0^+)\sim-\frac{1}{\ln|\omega|}$
[see Eq.~(\ref{eq:tta0w0})], and therefore the low temperature properties 
are still expected to be characterized by the strong coupling fixed point
as will be shown in Sect.~\ref{sec:nrg}.

The topology of the Fermi surface does not enter the present arguments at
any point, only the non-analyticity in the non-interacting 
DOS does. The latter derives from flat parts in the dispersion relation, which in the case of the 
$2d$ square lattice are saddle points at the $X$ points. It is therefore clear, that 
no nesting property  of the Fermi surface, which is present in the square 
lattice at half-filling,
can be responsible for these effects.

%%%%%%%%%%%%%%%%%%%%%%%%%%%%%%%%%%%%%%%%%%%%%%%%%%%%%%%
\section{Square lattice}
\label{sec:square}

\subsection{Half-filling}
\label{sec:halffill}

In all calculations of this work the hopping matrix element $t$ is 
used as the
energy scale and energies are measured in units $2t=1$,
setting the effective half bandwidth $W=2dt=2$. 
In this section the next-nearest neighbor hopping is 
set to zero, $t'=0$. 
The  non-interacting DOS of this $2d$ square lattice has the same  
logarithmic singularity at $\o=0$  
as the model DOS of Eq.~(\ref{eq:hubToyDosCusp}) 
for  $\alpha=0$.

\begin{figure}[th]
  \includegraphics[width=80mm]{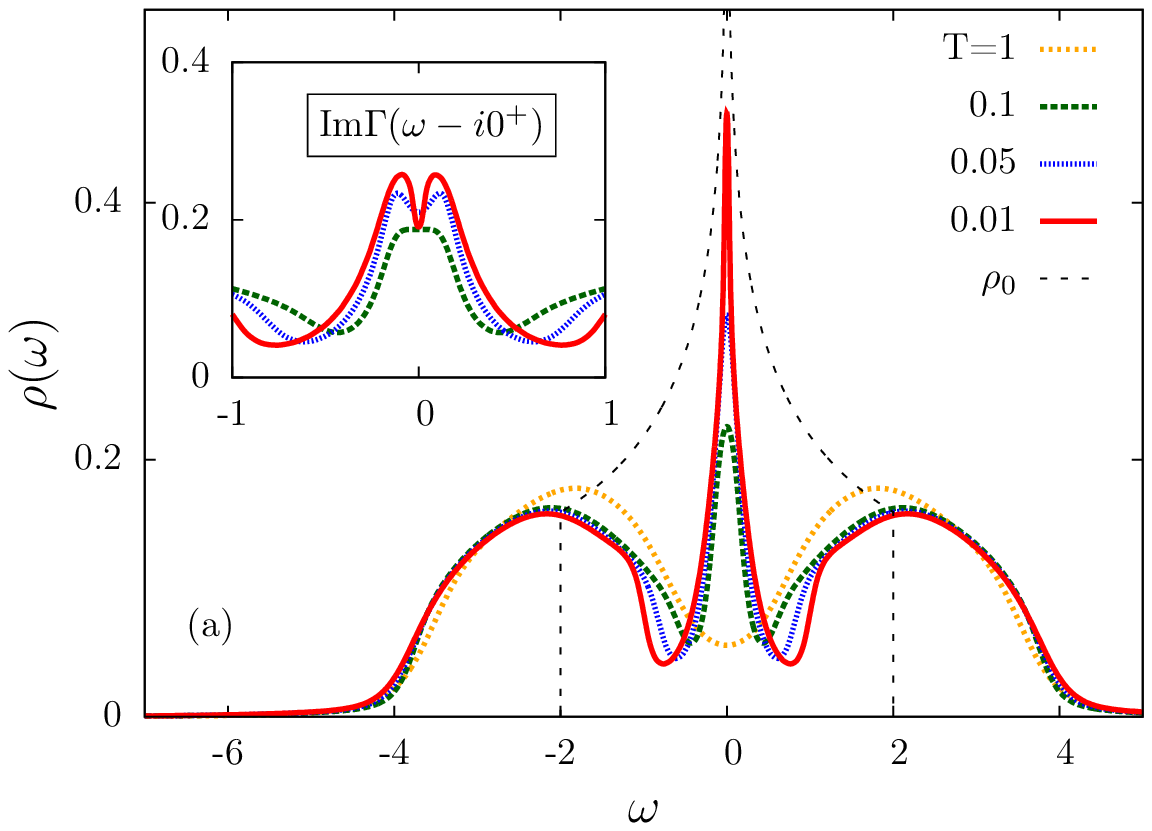} 
  \includegraphics[width=80mm]{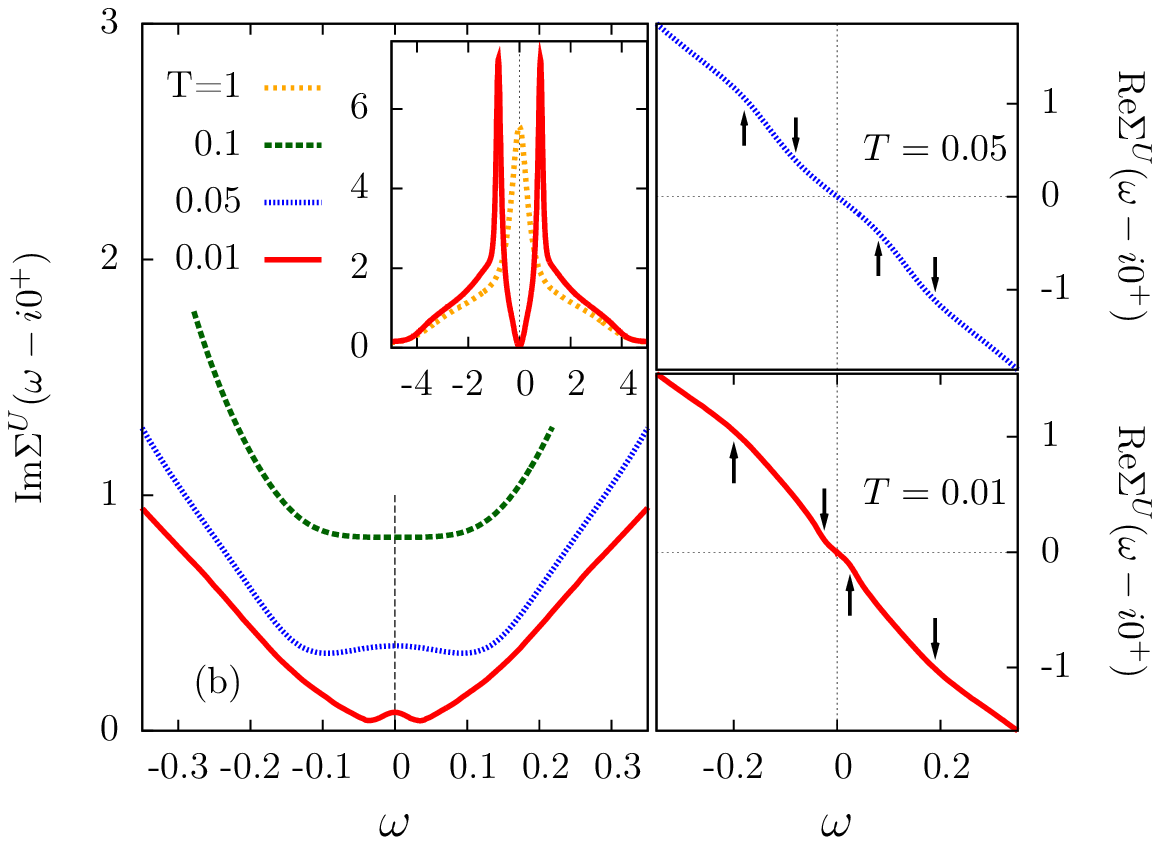}
 \caption{(a) Local spectral function $\rho(\o)$ for the symmetric 
   Hubbard model on a $2d$ square lattice  for Coulomb repulsion $U=4$ ($\e=-2$), 
   half bandwidth $W=2$ ($t=0.5$) and various temperatures  as functions
   of frequency within DMFT(ENCA).  The thin dashed 
   curve represents the non-interacting DOS $\rho_0(\o)$, and the inset shows the 
   imaginary part of the effective medium Im$\Gamma(\o-i0^+)$.
   (b) Imaginary (left) and real (right) part of the self-energy
   $\Sigma^U(\o-i0^+)$  for the same parameters as in (a) and $T$ as indicated.
   The inset of the left panel shows the imaginary part on a wider energy interval.
   The short arrows in the right graphs point at the approximate positions of the kinks.      
 } 
  \label{fig:hubb_2dsc_sym}
\end{figure}

Results of a DMFT(ENCA) calculation for the half-filled $n=1$
Hubbard model on a  square lattice are shown in Fig.~\ref{fig:hubb_2dsc_sym} 
for various temperatures. 
In part (a) the non-interacting  DOS as well as the fully interacting  
local spectral function $\rho(\o)=\frac{1}{\pi}\text{Im}G^{\text{(loc)}}(\o-i0^+)$ 
are displayed for $U=2W=4$.
The well separated Hubbard bands around the ionic level positions $\o\approx \pm U/2=\pm2$
are clearly visible.
Upon lowering the temperature the lattice version of the Kondo-effect
leads to the build-up of a pronounced many-body resonance at the 
Fermi level. 

The van Hove singularity in $\rho_0$ leads to the unusual minimum in the imaginary part
of the effective medium Im$\Gamma(\o-i0^+)$ 
[see inset of Fig~\ref{fig:hubb_2dsc_sym}(a)] as it was already found in the previous section.
In contrast to
Fig.~\ref{fig:hubToyDosTtilde} the 
minimum in Im$\Gamma(\o-i0^+)$ is not cusp-like at $\o=0$ but smeared out due to the  
imaginary part of the self-energy which is shown in part (b) of the figure. 
However, the non-analytic structure in the lattice scattering matrix, Eqs.~(\ref{eq:scatter})
and (\ref{eq:hubToyLinDos0}), is sufficient to produce this minimum at low temperatures.

The self-consistency of DMFT leads to an accompanying maximum in 
Im$\Sigma^U(\o-i0^+)$ at $\o=0$, which is in striking contrast to Fermi-liquid 
behavior Im$\Sigma^U(\o-i0^+)\sim \o^2+\pi^2T^2$ for small $\o$ and $T$. 
The position of the two minima emerging at finite frequencies next to the central maximum is
moved toward the Fermi level with decreasing temperature so that the anomalous
region shrinks. 

As a side remark we add here, that we observe 
additional kinks in the real part of the 
self-energy as shown in Fig.~\ref{fig:hubb_2dsc_sym}(b), which are
associated with this non-Fermi liquid behavior.
At energies $\o\approx \pm0.2$ the known kinks arising 
from the coupling of the quasiparticles to
local spin fluctuations are observable.\cite{byczukKinks09,raasCollectiveMode09}
The additional kinks are clearly visible at lower energies.
In contrast to the kinks at higher energy,  their position
is temperature dependent and  moves 
toward the Fermi level with decreasing $T$.

Anomalous maxima in Im$\Sigma^U$ are also encountered within the 
functional renormalization group approach to the Hubbard model
at finite temperatures.\cite{kataninQPRG04,roheNFL05} Crucial for these
anomalies to occur is the renormalization of
the two-particle interaction vertex. Within DMFT
all \textit{local} $n$-particle vertices and their renormalizations 
are included.

\begin{figure}[t!]
  \includegraphics[width=80mm]{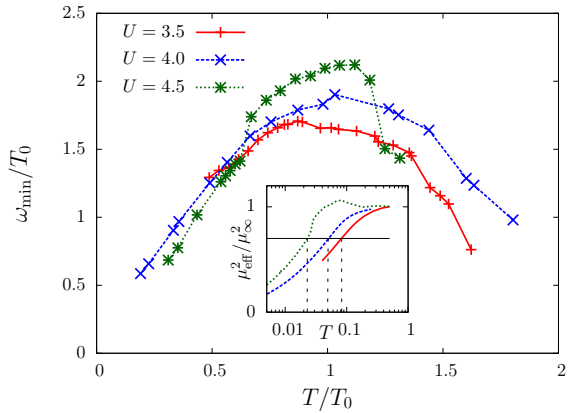}
  \caption{
    Position of the minimum $\o_{\text{min}}$ in Im$\Sigma^U$ in 
    units of the low-energy scale $T_0$
    as a function of 
    the rescaled temperature $T/T_0$  for three different Coulomb repulsions. 
    The inset shows the screened local magnetic moment normalized to its high temperature
    value $\mu^2_\infty$ as a function of temperature
    for the same values of the Coulomb repulsions.
    The horizontal line 
    indicates the value where $\mu^2_{eff}$ reaches $70\%$ of $\mu^2_\infty$
    and the vertical dashed lines indicate the low-energy scales: $T_0(U=3.5)\approx 0.082$,
    $T_0(U=4)\approx 0.051$, and $T_0(U=4.5)\approx 0.023$.
  } 
  \label{fig:hub_2dsc_scale}
\end{figure}

We extract the low-energy scale $T_0$ as the temperature 
where  the  effective screened
local moment as calculated from the local magnetic 
susceptibility via\cite{schmittSus09}
$\mu_{\text{eff}}^2=T\chi^{\text{mag}}_{\text{loc}}$
is reduced to $70\%$ of its high temperature
value $\mu^2_\infty$ (see inset of Fig.~\ref{fig:hub_2dsc_scale}).
This yields the equivalent result to the width of the many-body resonance 
at the Fermi level  in the momentum resolved spectral function Im$G(t_{\k_\text{F}},\o-i0^+)/\pi$
(not shown).

The  temperature evolution of the anomalous double-well structure is governed by
the low-energy scale as revealed by  Fig.~\ref{fig:hub_2dsc_scale}.
There,  the position of the anomalous minimum $\o_{\text{min}}$  in
Im$\Sigma^U$ as a function of temperature
for three values of the Coulomb repulsion is shown (the self-energies
as well as the minima at $\pm \o_{\text{min}}$ are symmetric around $\o=0$ for half-filling).
For all $U$, $\o_{\text{min}}/T_0$ exhibits a dome-shaped curve with
both, maximum position and height determined by $T_0$.
The physical origin can therefore
be directly linked to the emergence of the low-energy quasiparticles.
At high temperatures the system is dominated by the incoherent charge and spin excitations 
and the van Hove singularity is concealed. 
At temperatures of the order of $T_0$ 
the lattice version of the Kondo effect leads to the screening of local magnetic 
moments.
The break-up of screened magnetic moments gives rise to long-lived 
low-energy quasiparticle excitations which manifest themselves in the 
many-body resonance  at the Fermi level as observed in 
the spectral function of  e.g.\ Fig.~\ref{fig:hubb_2dsc_sym}(a).
%the low energy quasiparticles start to form 
But their evolution into the coherent Fermi liquid quasiparticle is 
disturbed by the enhanced scattering due to the van Hove singularity. 
This leads to the maximum in Im$\Sigma^U$ at the Fermi level.
Further lowering  the temperature  the phase space volume for 
quasiparticle scattering shrinks and so does  the extension of the 
maximum.

%%%%%%%%%%%%%%%%%%%%%%%%%%%%%%%%%%%%%%%%%%%%%%%%%%%%%%%%%%%%%%%%%%%%%%%%%%%%%%%%%%%%%%%%%%%%%%%%%%%%%%%%%%%%%%%
\subsection{Finite doping}
\label{sec:doped}
\begin{figure}[t!]
    \includegraphics[width=80mm]{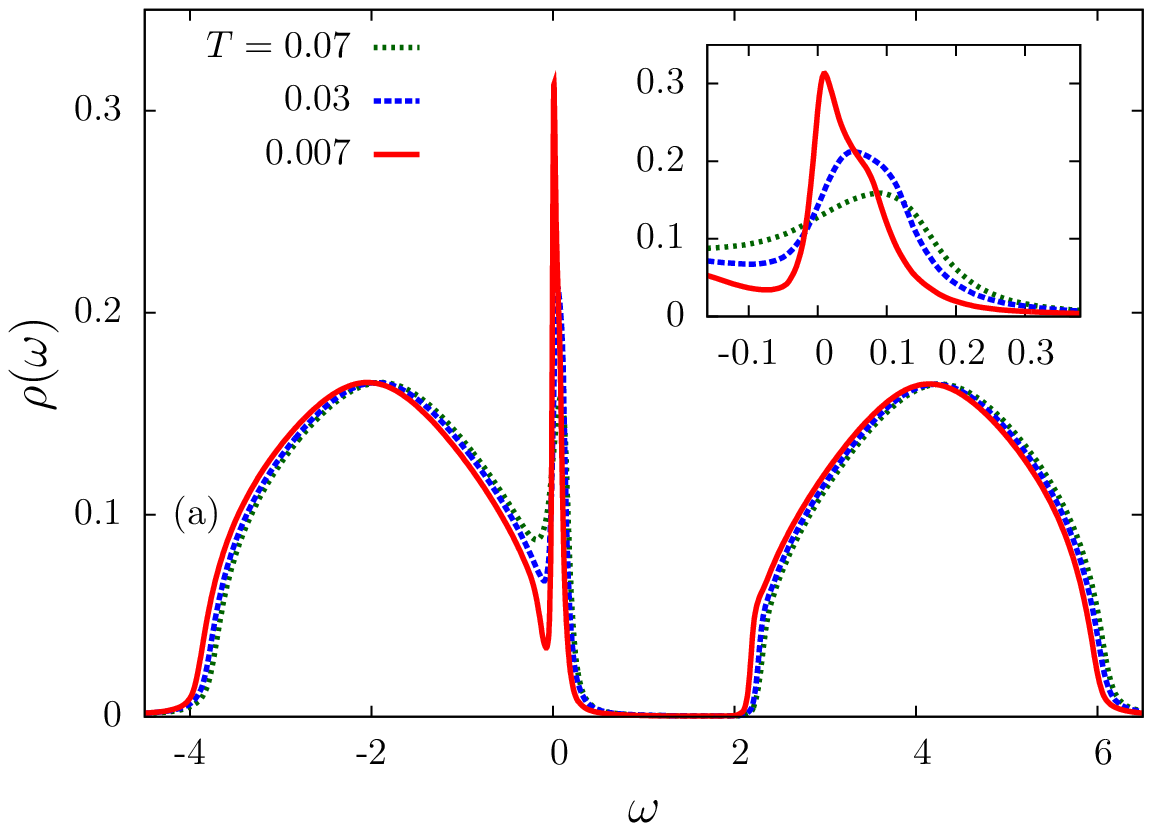}
    \includegraphics[width=80mm]{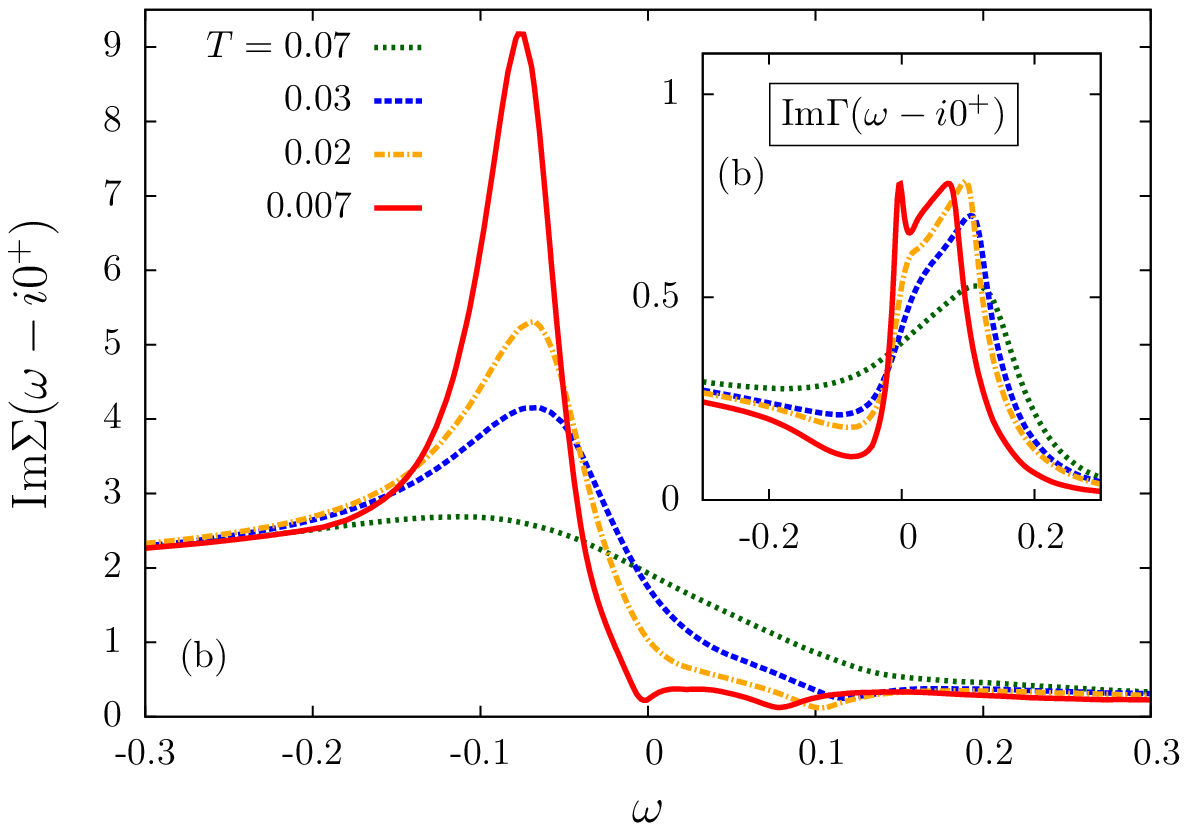}
  \caption{(a) Local DOS for the asymmetric Hubbard model on a $2d$ square lattice
    with $U=6$ and filling $n=0.98$ for various temperatures obtained with 
    the DMFT(ENCA).  The inset shows the low-energy
    region.
    (b) Imaginary part of the self-energy around the Fermi level for the same parameters as in (a) and temperatures as
    indicated. The inset shows the imaginary part of the effective medium in the low-energy 
    region.
  } 
  \label{fig:hubb_gf_2dsc_asym}
\end{figure}

The van Hove singularity 
is shifted away from the Fermi energy $\o=0$
when doping the system, $n=1-\delta$.
The corresponding anomalous minimum in the effective medium Im$\Gamma$ and the maximum 
in Im$\Sigma^U$ are too moved to finite energies.
This can be seen in Fig.~\ref{fig:hubb_gf_2dsc_asym}, where 
the spectral function, the self-energy, and the effective medium are shown for the Hubbard model
on a $2d$ square lattice ($t'=0$) with $U=6$ and $\delta=0.02$ (filling $n=0.98$).

The quasiparticle peak around $\o=0$ in the spectral function, as displayed in the inset 
of Fig.~\ref{fig:hubb_gf_2dsc_asym}(a), acquires a pronounced asymmetry due to the double-well
structure in Im$\Sigma^U$. The shoulder in the curve for the lowest temperature might
even be interpreted as a precursor of a pseudogap.

The flat parts in the dispersion relation
are located energies away from the Fermi level and the dispersion relation can
again be approximated linearly in a vary narrow region around the Fermi surface. 
In this region the usual  arguments of microscopic Fermi liquid theory hold and the 
system forms a regular Fermi liquid at very low temperatures. This is observed in
the self-energy shown in Fig.~\ref{fig:hubb_gf_2dsc_asym}(b) where a quadratic minimum 
forms at $\o=0$  at $T=0.007$. 
However, the Fermi liquid parameters will be strongly renormalized as compared to a situation
without the van Hove singularity.

%%%%%%%%%%%%%%%%%%%%%%%%%%%%%%%%%%%%%%%%%%%%%%%%%%%%%%%%%%%%%%%%%%
\subsection{Self-consistent treatment with a model DOS}
\label{sec:modelDOSDMFT}
\begin{figure}[t]
  \centering
  \includegraphics[width=40mm]{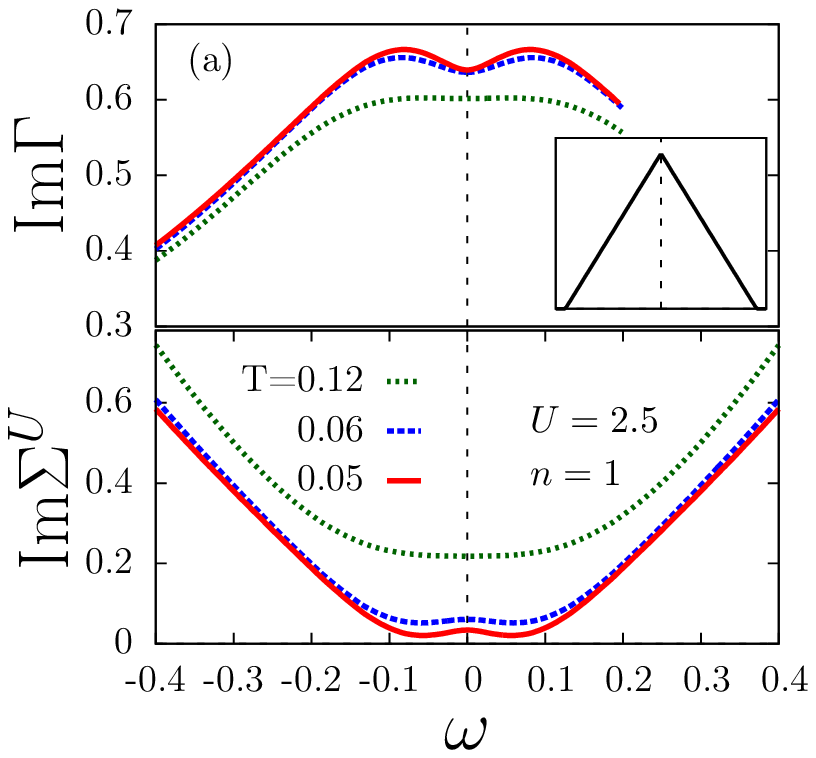}\includegraphics[width=40mm]{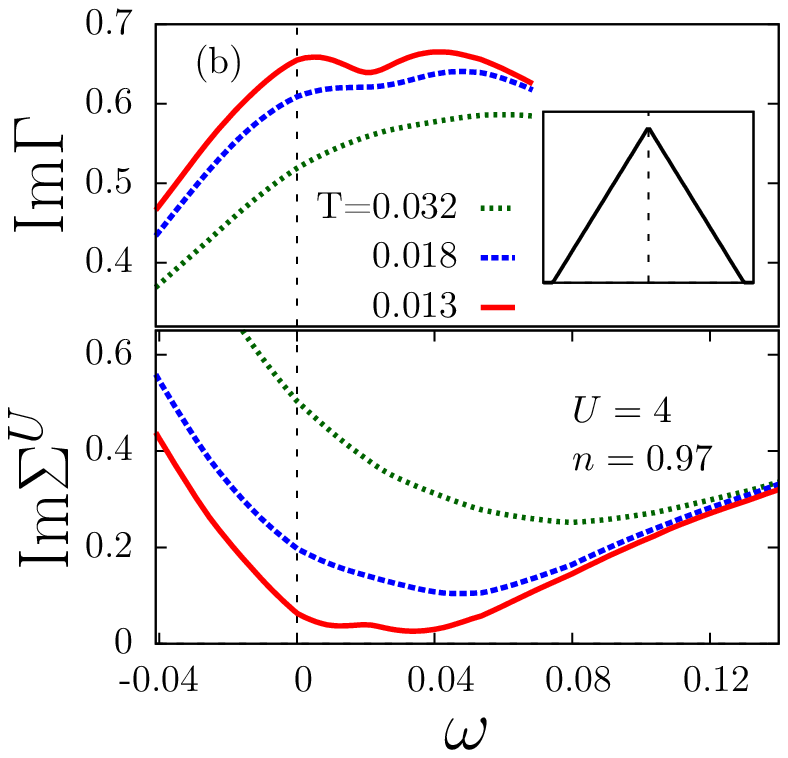}\\
  \includegraphics[width=40mm]{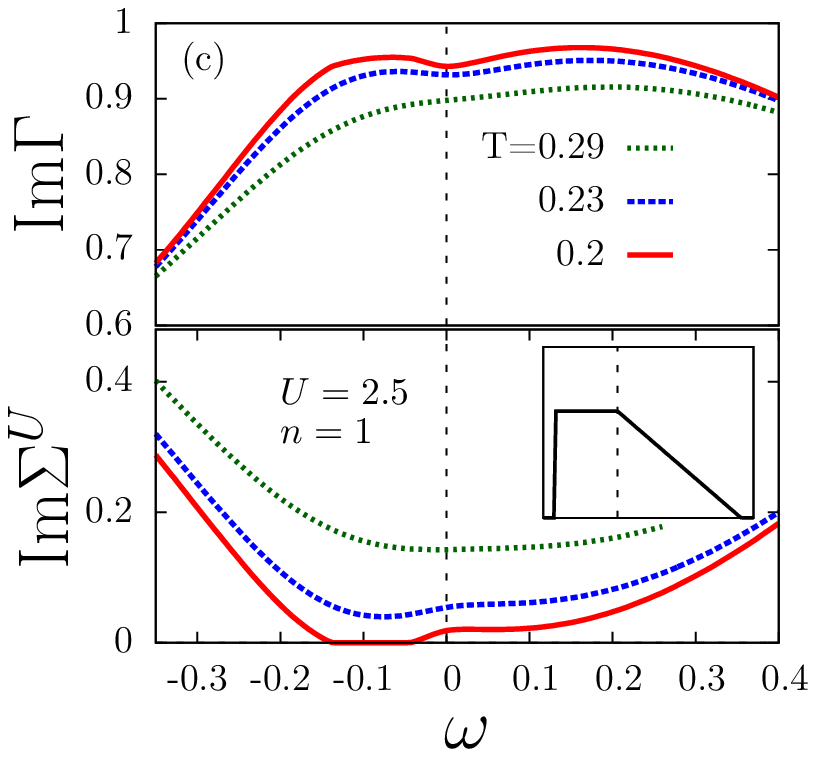}\includegraphics[width=40mm]{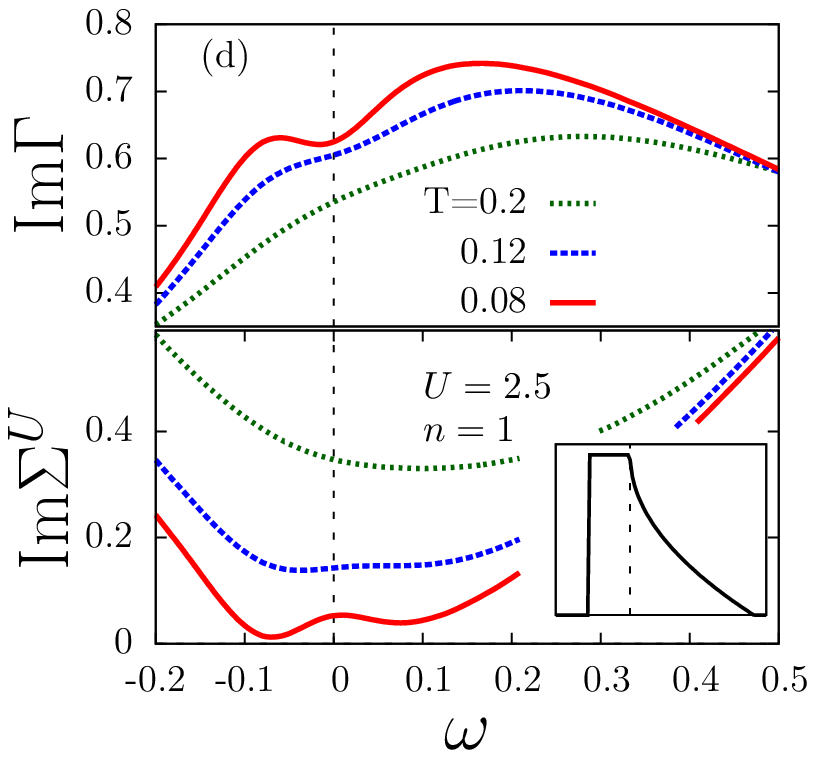}\\
  \caption{The imaginary parts of the effective medium (upper graphs) and self-energy (lower graphs) 
    of the Hubbard model within DMFT(ENCA) for different non-interacting DOS, which are shown
    in the insets. The vertical dashed line indicates the position of the Fermi level.
    The parameters are indicated in the plots.
  } 
  \label{fig:hubRealDosRunsTTSG}
\end{figure}

No true divergence in the non-interacting DOS  is necessary for the 
anomalies described above  to occur. A non-analytic 
cusp as in the model DOS of Eq.~(\ref{eq:hubToyDosCusp}) for $0<\alpha<2$  is sufficient. 
This is illustrated in Fig.~\ref{fig:hubRealDosRunsTTSG}
where  fully self-consistent DMFT calculations for several model DOS
are  considered.  The first two plots (a) and (b) are 
for $\rho^{\text{cusp}}_\alpha(\o)$  with $\alpha=1$ at and 
away from half-filling, respectively. Figures  (c)
and (d) are for asymmetric DOS which are constant below the Fermi level
and have linear (c) and  a square-root (d) cusp  at $\o=0$.
The imaginary parts of the effective media and self-energies show the characteristic 
minima and maxima induced by the non-analyticities in $\rho_0$.
For the half-filled cases (a), (c) and (d) the
anomalies are situated at the Fermi level, while for finite hole doping as in (b) 
these are moved to positive energies.

%%%%%%%%%%%%%%%%%%%%%%%%%%%%%%%%%%%%%%%%%%%%%%%%%%%%%%%%%%%%%%%%%%
\subsection{Zero Temperature}
\label{sec:nrg}

\begin{figure}[t]
     \includegraphics[clip=true,width=80mm]{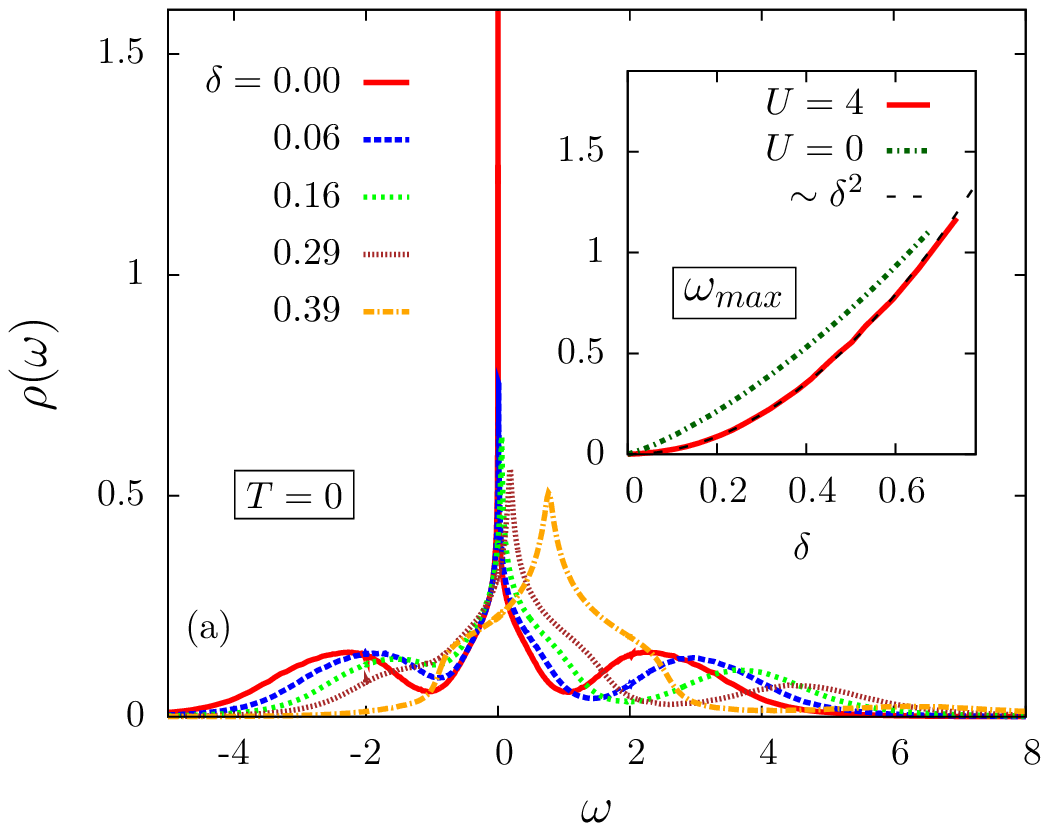}
   \includegraphics[clip=true,width=80mm]{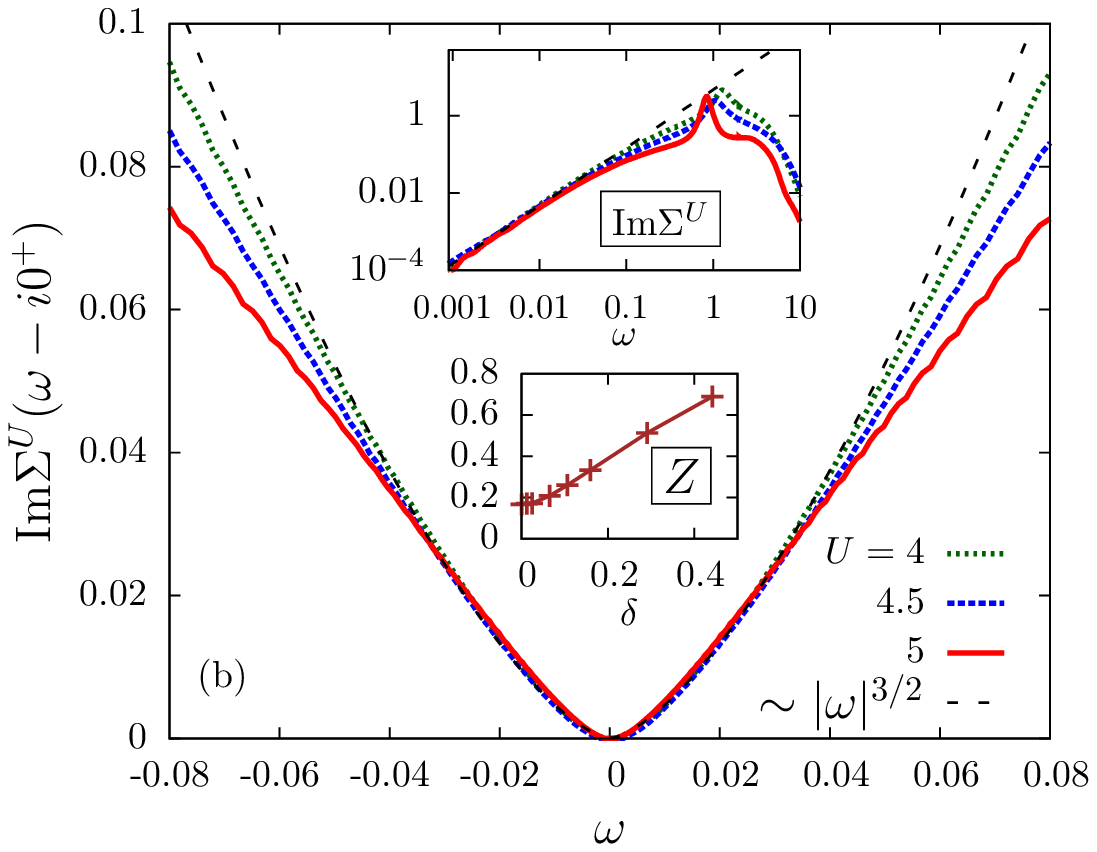}
   \caption{(a) DMFT(NRG) spectral function at $T=0$ for the  Hubbard model on a $2d$ square lattice for $U=4$ and 
     varying hole doping. The inset shows the position of the maximum as a function of doping for $U=4$ and $U=0$.
     The dashed line is a fit $\o_{\text{max}}=2.2\delta^2$. (b) Imaginary part of the self-energy at half-filling 
     $\delta=0$  for three values of the Coulomb repulsion $U$. The curves for $U=4.5$ and $U=5$ were rescaled 
     in order to lie on top of the  $U=4$ curves at low $\o$. The dashed line is a fit with a function $a \,|\o|^{3/2}$. The upper inset shows the low energy behavior in a double-logarithmic plot.
     The lower inset shows the
     quasiparticle weight for $U=4$ as a function of doping.
  } 
  \label{fig:hub-nrg_2d_sc_sym}
\end{figure}

Figure~\ref{fig:hub-nrg_2d_sc_sym}(a) displays  the spectral function
for the Hubbard model on a square lattice with $U=4$ at zero temperature and
for different doping $\delta$. These spectra were obtained with the NRG\cite{petersNewNRG06} 
as impurity solver.

As the hole-doping is increased, the lower Hubbard band moves toward the Fermi level
and eventually merges with the quasiparticle peak.
For very large doping the lower Hubbard band is moved above the Fermi level
and $\rho(\omega)$ resembles the non-interacting DOS
as  the system becomes effectively non-interacting. 

The position of the quasiparticle peak is attracted to the Fermi level up to
considerable values of doping (roughly $\delta\lesssim 0.2$).
This pinning becomes more visible in the inset, where the position $\o_{\text{max}}$
of the maximum in $\rho(\o)$ is plotted as a function of doping $\delta$ for $U=4$ and $U=0$.
In the interacting case $\o_{\text{max}}$ depends quadratically on the doping, $\o_{\text{max}}\sim\delta^2$, and is
considerably reduced compared to $U=0$. 
The many-body correlations renormalize the van Hove singularity toward the Fermi level especially 
at small doping $\delta\lesssim 0.2$.

The imaginary part of the self-energy shown in Fig.~\ref{fig:hub-nrg_2d_sc_sym}(b) does not 
display any anomalous double-well structure.
The logarithmic van Hove singularity in $\rho_0$ is a rather weak divergence
and at zero temperature the usual arguments leading to a vanishing self-energy at the Fermi 
level apply.\cite{luttinger:AnalyticPropGf61} 
However, Im$\Sigma^U$ shows an anomalous frequency dependence at low energies,
as it grows like Im$\Sigma^U(\o-i0^+)\sim|\o|^{3/2}$ which 
was also previously found in perturbation theory.\cite{gopalanSaddle92,hlubninaHotSpot95}
\footnote{The exponent $3/2$ is found in perturbation theory for quasiparticles
at the Fermi level, but not at the van Hove singularity, which scatter from
quasiparticles at the van Hove singularities. For quasiparticles 
at the van Hove singularity  Im$\Sigma^U(\o-i0^+)\sim |\o|$ and scattering processes
not involving the saddle points produce the usual $2d$ Fermi liquid result
Im$\Sigma^U(\o-i0^+)\sim \o^2\ln 1/|\o|$.
DMFT does not resolve the different momentum dependencies in the interaction
channels
and averages over the whole Brillouin zone. Due to its large phase space, the
contribution $\sim|\o|^{3/2}$ dominates this average and is therefore
observed in the DMFT. This is confirmed by 
fitting the imaginary part with a function
$A|\o|^{3/2}+B|\o|$ which yields a very small linear contribution 
$B/A\approx 1-2 \%$ without changing the range of validity of the fit.
The usual $2d$ Fermi contribution $\sim \o^2\ln 1/|\o|$ vanishes too fast
for $\o\to 0$  to make a significant contribution at low energies.
} 
The upper inset reveals this low energy behavior more clearly in a double logarithmic plot.

The quasiparticle peak at the Fermi level for half-filling $\delta=0$ 
shows a logarithmic divergence as was already  observed recently.\cite{zitkoVHSDMFT09} 
But in contrast to what was stated in Reference~\onlinecite{zitkoVHSDMFT09},
the quasiparticle  weight 
\begin{align}
\label{eq:qpzT0}
  Z&=\frac{1}{1-\frac{\partial\text{Re}\Sigma^U }{\partial \o}\big|_{\o=0}}
\end{align}
does not vanish at half-filling in our calculation,
as can be seen in the inset of Fig.~\ref{fig:hub-nrg_2d_sc_sym}(b). 
(The quasiparticle weight in Fig.~3 of Ref.~\onlinecite{zitkoVHSDMFT09}
also seems to extrapolate to a finite value at half-filling
but a contradicting statement is made in the text of that 
reference.) 
The slope of the real part of the self-energy does consequently not diverge at the Fermi level. 
This is in accord with the low frequency dependency of the imaginary part $\sim |\o|^{3/2}$
from which
the  real part is obtained 
via Kramers-Kronig relation,
Re$\Sigma^U(\omega-i0^+)\sim \omega[\sqrt{\omega }(\text{acoth}\sqrt{\omega }+\text{arccot}\sqrt{\omega })-2]$. 

The system is therefore well characterized by a generalized  Fermi 
liquid\cite{gonzalesSIAMPseudogap98,loganPseudogapSIAM00,glossopPseudogap00}
where Im$\Sigma^U$ shows some anomalous frequency dependence, but still vanishes at the Fermi level
and the quasiparticle weight remains finite.

%%%%%%%%%%%%%%%%%%%%%%%%%%%%%%%%%%%%%%%%%%%%%%%%%%%%%%%
\section{Next-nearest neighbor hopping}
\label{sec:ttp}
%%%%%%%%%%%%%%%%%%%%%%%%%%%%%%%%%%%%%%%%%%%%%%%%%%%%%%%
\subsection{Spectral function and self-energy  }
\label{sec:gfSig}

\begin{figure}[t]
\includegraphics[width=45mm]{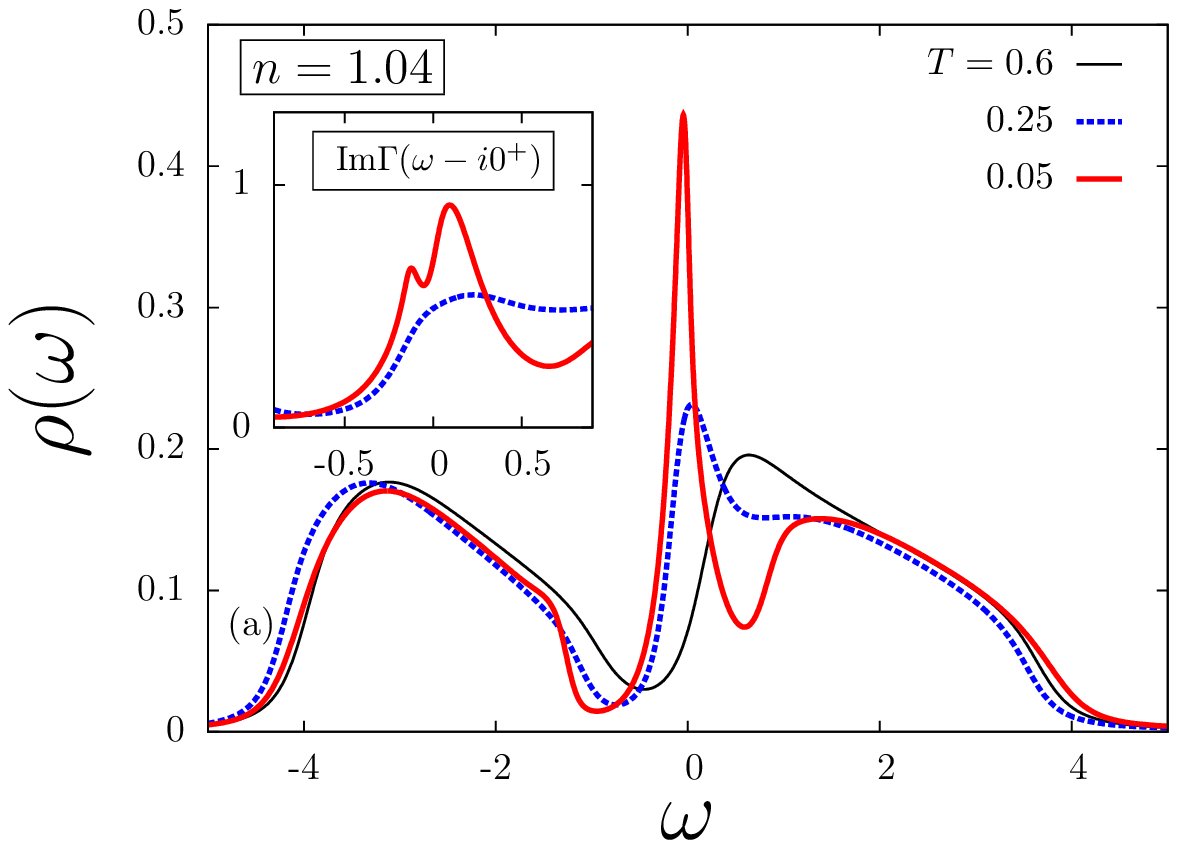}\includegraphics[width=45mm]{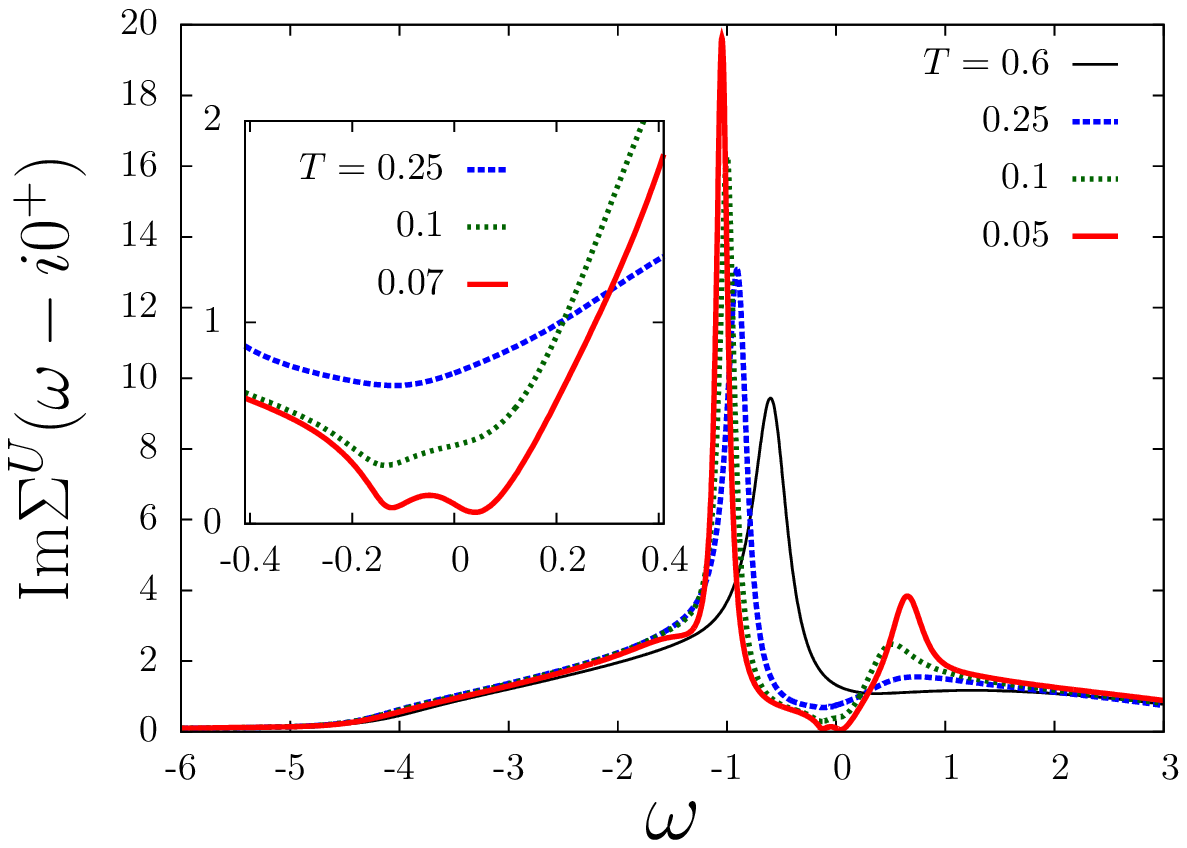}
\includegraphics[width=45mm]{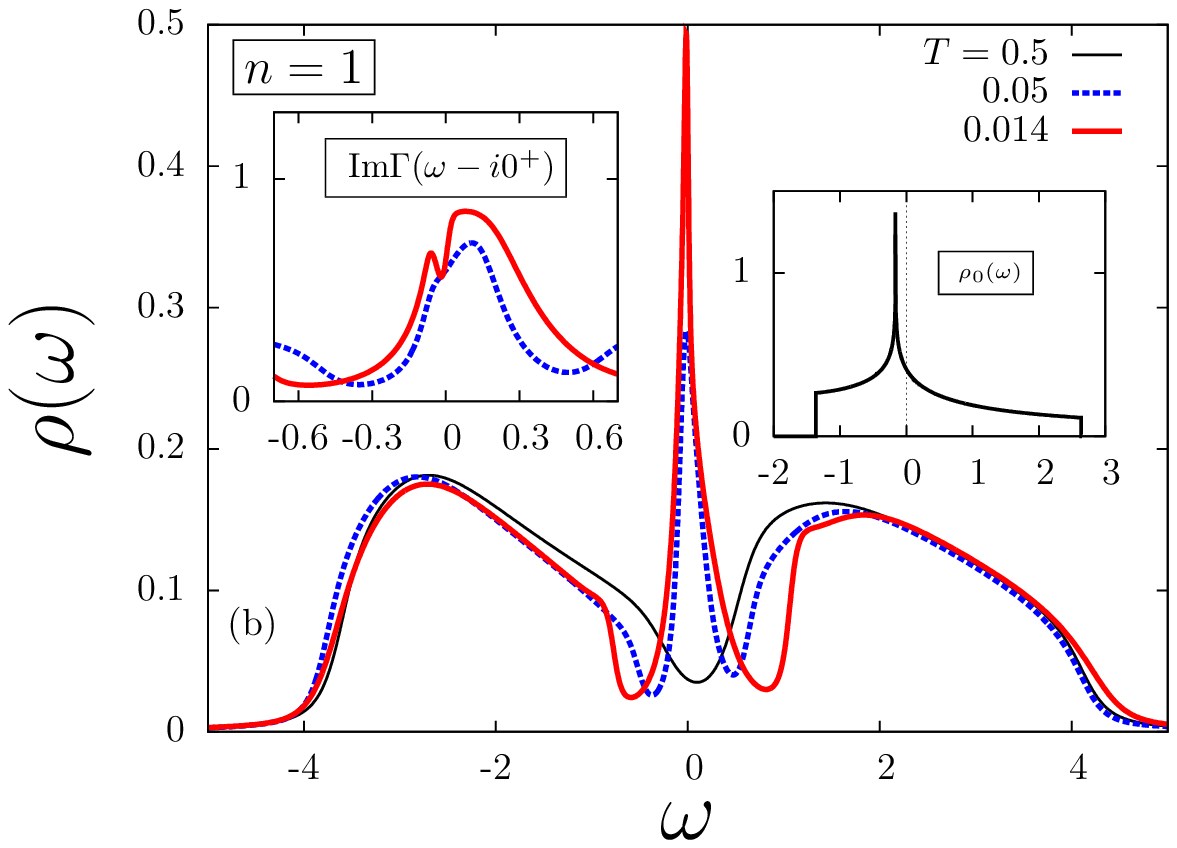}\includegraphics[width=45mm]{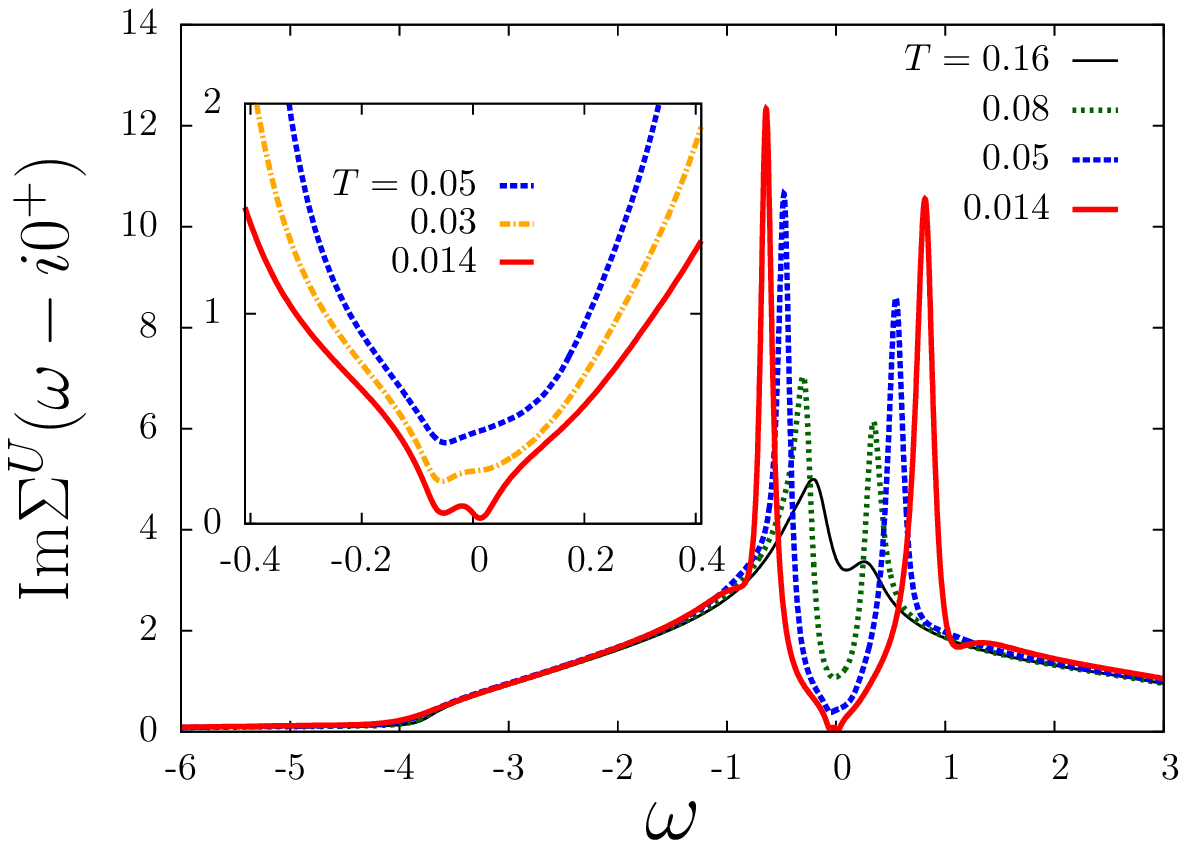}
 \includegraphics[width=45mm]{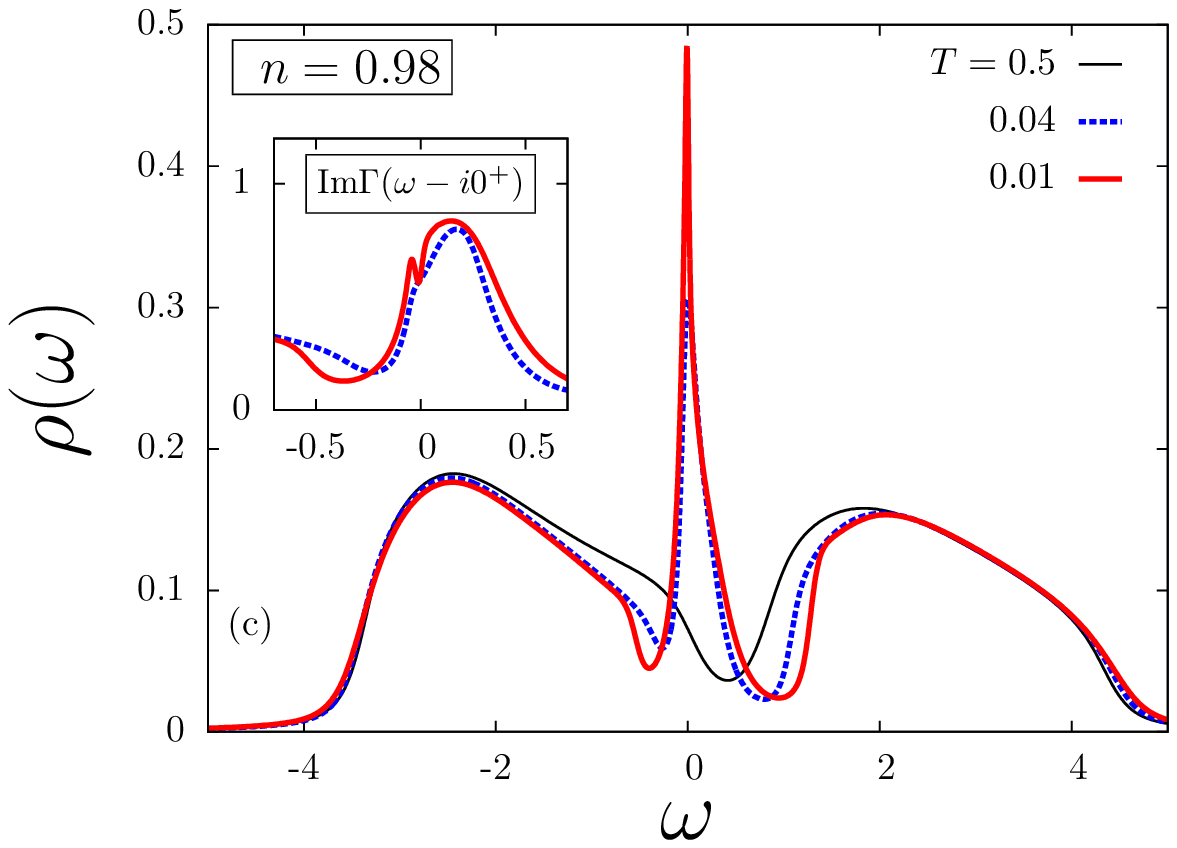}\includegraphics[width=45mm]{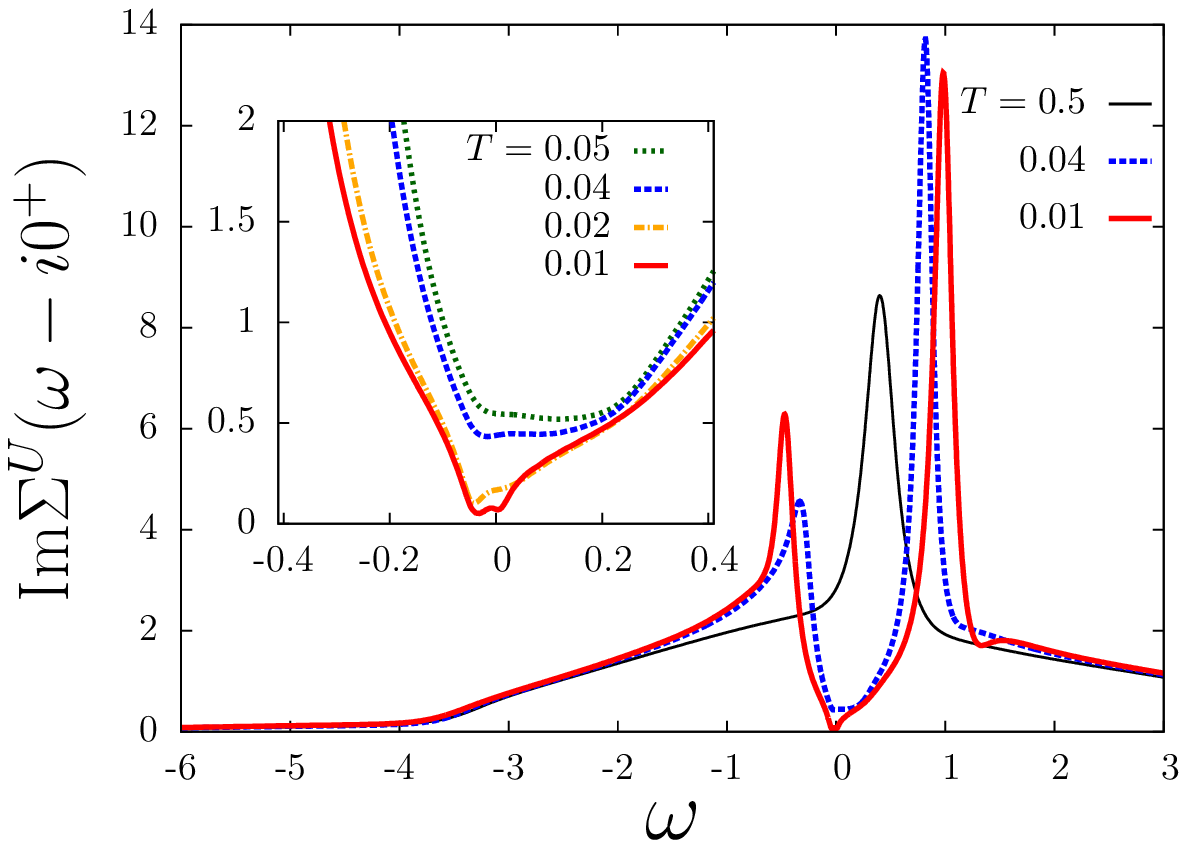}
 \includegraphics[width=45mm]{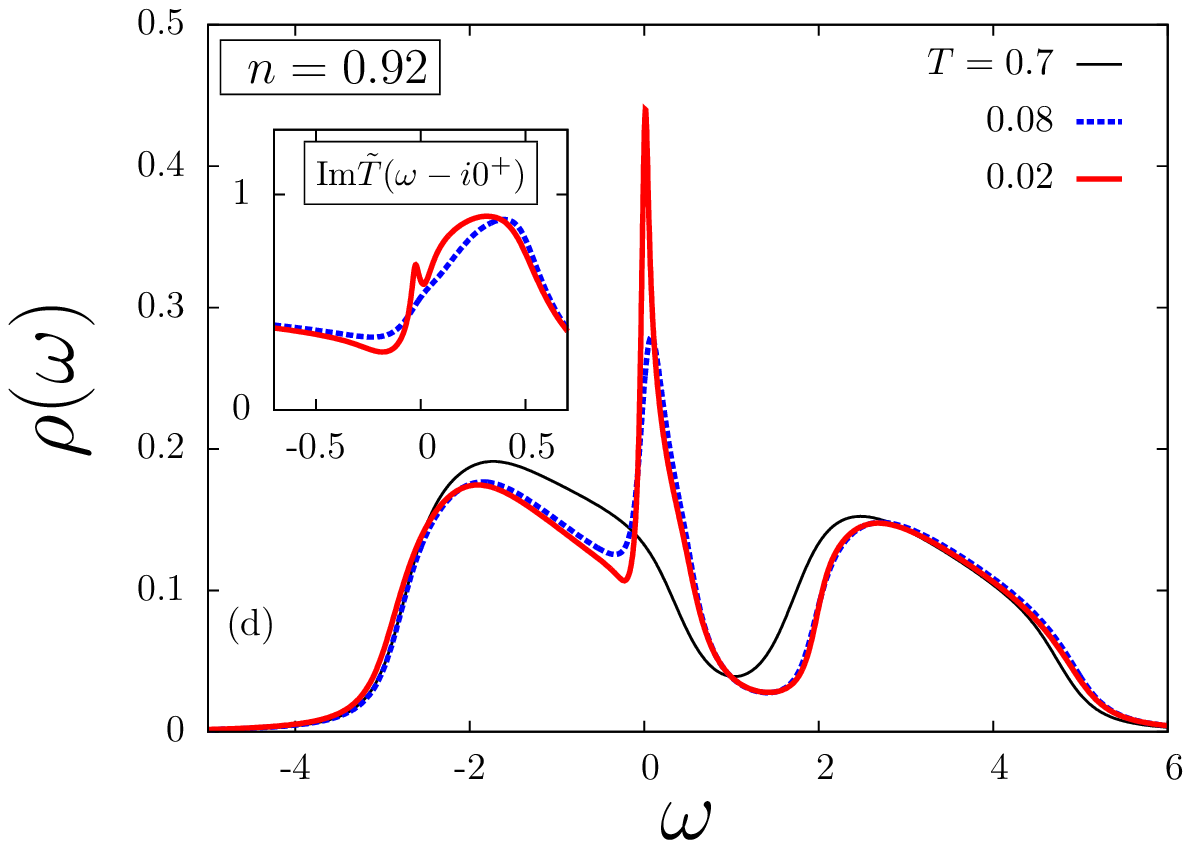}\includegraphics[width=45mm]{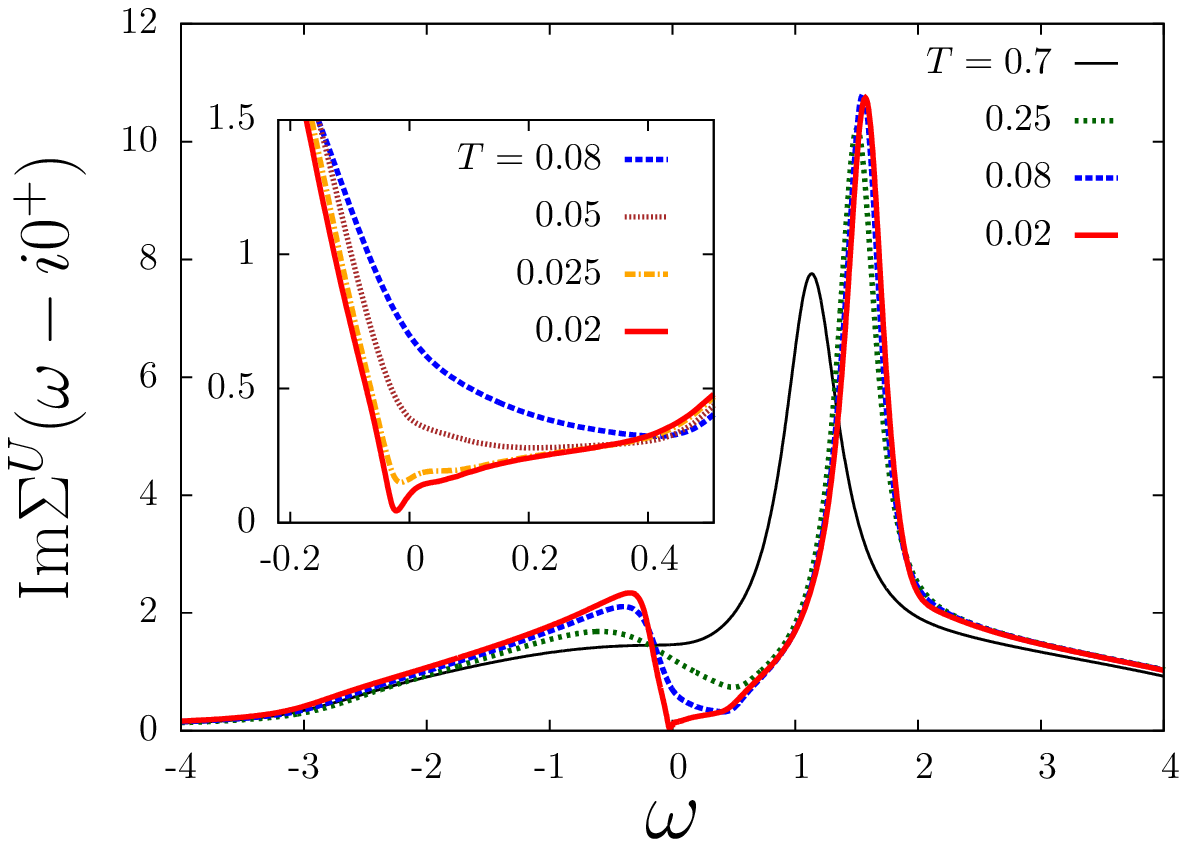}
  \caption{DMFT(ENCA) spectral functions $\rho(\o)$ (left panels) and  imaginary parts of the 
    self-energy $\Sigma^U(\o-i0^+)$ (right panels) for different filling $n$ 
    for the Hubbard model on the square lattice with additional next-nearest neighbor hopping $t'=-0.2t=-0.1$
    and $U=4.25$.  The insets in the left panels show the imaginary part of the effective medium
    while the insets on the right show a close-up of the low-energy region of Im$\Sigma^U$.
    The non-interacting DOS is shown in the inset for the half-filled case (b).  
  } 
  \label{fig:hubb_2dscnnn_asym}
\end{figure}

In this section we  extend the study to an additional
next-nearest neighbor hopping $t'=-0.2t$.
The local spectral functions and self-energies displayed in Fig.~\ref{fig:hubb_2dscnnn_asym}
for  $U=4.25$, various fillings and temperatures show
pronounced particle-hole asymmetries.

At low temperatures the anomalous minima in Im$\Gamma$ and corresponding maxima in 
the self-energy Im$\Sigma^U$ can be observed slightly below the  Fermi level
 for filling $n=1.04$, $n=1$, and $n=0.98$.
This correlates with 
the  logarithmic van Hove singularity in $\rho_0$ shifted below the Fermi level, 
as displayed in the right inset of 
Fig.~\ref{fig:hubb_2dscnnn_asym}(b). 
For decreasing filling (increasing hole doping) the van Hove singularity is moved
toward the Fermi level and the anomalous structures in Im$\Sigma^U$ and Im$\Gamma$
narrow.  

For the largest hole doping $n=0.92$ the imaginary part of the self-energy exhibits
only one narrow minimum at very low temperature. 
Within DMFT the Fermi surface does not change upon increasing the interaction $U$,\cite{muellerHartmannDInfty89-2} 
due to the momentum independence of the self-energy. The filling at 
which the van Hove singularity moves across the Fermi level can be calculated 
with the non-interacting DOS to be $n\approx 0.82$, which is much less than $n=0.92$
Therefore a double-well structure would be expected but is not observed. 
The reason is found in the lower Hubbard band  reaching  
the Fermi level. A growing number of empty lattice sites are 
created and disturb the quasiparticle formation. Additional quasiparticle 
scattering results and conceals the effect of the van Hove singularity. 

Due to the pinning of the singularity to the neighborhood of the 
Fermi level its influence is strong  
for all values of filling.

The anomalies correlate with the low-energy scale and
occur at temperatures following the trend given by
$T_0$ as function of filling $n$.
$T_0$ reaches a minimum
for $n \approx 0.98$ to $1$ where 
correlations are strongest and the system is 
closest to the Mott  insulator.

%%%%%%%%%%%%%%%%%%%%%%%%%%%%%%%%%%%%%%%%%%%%%%%%%%%%%%%
\subsection{Transport}
\label{sec:trans}

\begin{figure}[t]
  \includegraphics[width=80mm]{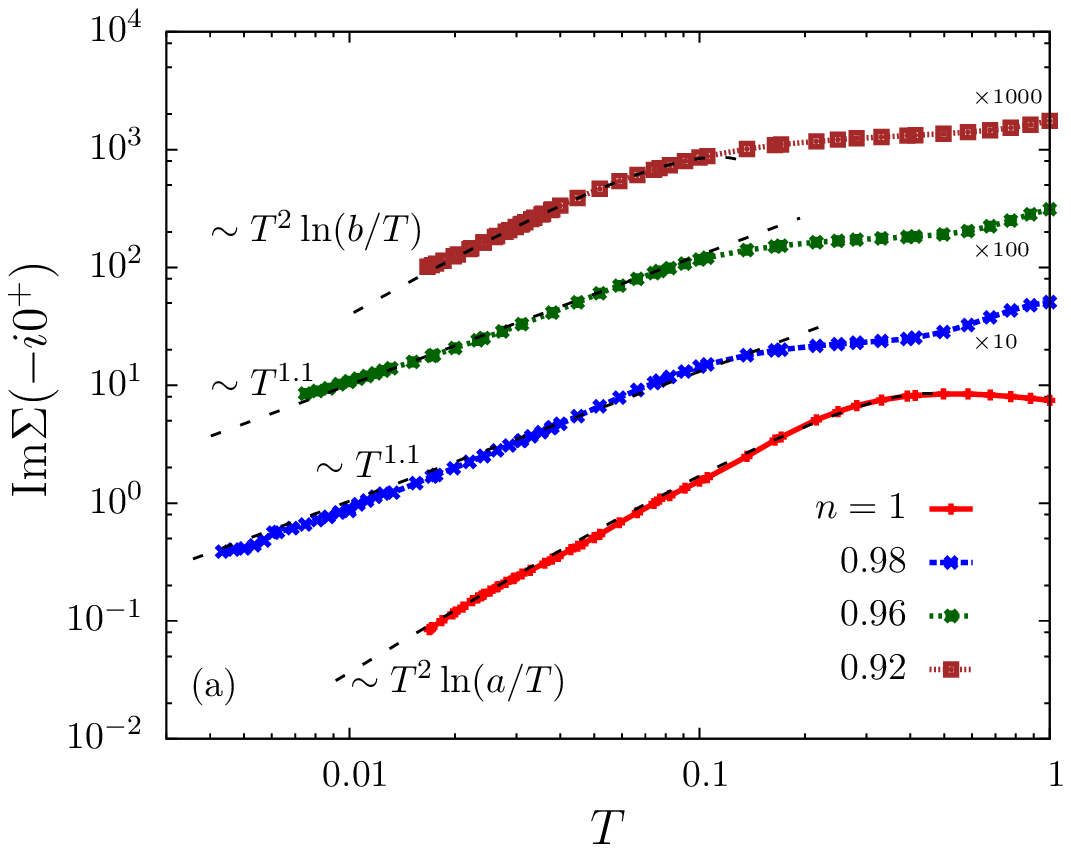}
  \includegraphics[width=80mm]{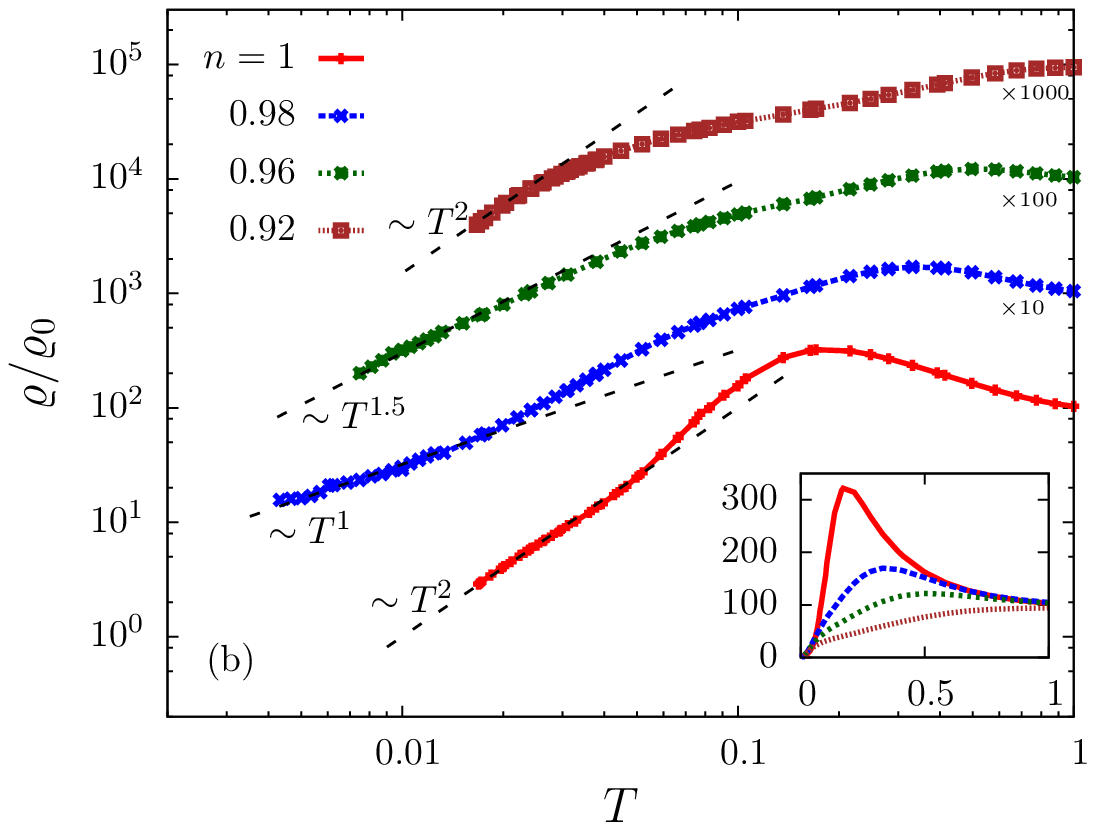}
  \caption{(a) Static quasiparticle scattering rate Im$\Sigma^U(-i0^+)$ 
    for the Hubbard with $U=4.25$ on a square lattice with next-nearest neighbor hopping 
    $t'=-0.2t$ as function of temperature in log-log plot. The curves for different filling 
    are multiplied by the factors  indicated on the right.
    (b) Resistivity $\varrho$  (normalized to $\varrho_0$) for the same parameters as in (a). The inset shows $\varrho$
    without the scaling factors and in a non-logarithmic plot. 
    The dashed lines represent fits with the indicated functions  in both plots.
  } 
  \label{fig:hubb_2dscnnn_asymTrans}
\end{figure}

The anomalous behavior found in the self-energy affects the temperature dependence 
of transport properties.  The static quasiparticle scattering rate at the 
Fermi level Im$\Sigma^U(-i0^+)$ is shown in Fig.~\ref{fig:hubb_2dscnnn_asymTrans}(a)
as a function of temperature for different filling. 

For a $2d$ Fermi liquid the expected temperature dependence 
of the scattering rate is given by
Im$\Sigma^U(-i0^+,T)\sim T^2\ln(1/T)$.
This behavior is found for low temperatures at half-filling [lowest 
curve in Fig.~\ref{fig:hubb_2dscnnn_asymTrans}(a)]
and above, i.e.\ $n\gtrsim 1$ (not shown). 
Hole-doping induces qualitative 
changes in the low temperature behavior since the van Hove singularity is
moved toward the Fermi level.
For filling $n=0.98$ and $n=0.96$  power-law behavior with an exponent close
to one is observed, 
Im$\Sigma^U(-i0^+,T)\sim T^{1.1}$.
For large hole doping, $n=0.92$, the scattering rate is again 
found to be in accord with usual Fermi liquid variation, $\sim T^2\ln(1/T)$.
This is expected to hold for even larger hole-doping $n\lesssim 0.92$, 
but could not be checked 
within the present approach due to restriction of the ENCA to relatively 
high temperatures at large doping.

We want to emphasis  that the van Hove singularity 
induces qualitative changes in the asymptotic low temperature behavior of the scattering 
rate when moved closer to the Fermi surface by doping. 
Im$\Sigma^U(-i0^+,T)$  decreases much 
slower with temperature for fillings around  $n\approx 0.96$ when the 
double-well structure
is found  in the vicinity of the Fermi level. 

With the self-energy at hand one can also calculate transport coefficients.  
No vertex corrections occur in DMFT due to the momentum independence of the self-energy
and two-particle vertex.\cite{khuranaResistivityHM90,muellerHartmannDInfty89,schweitzerHeavyFermionTransport91}
The current-current correlation function is completely determined by the particle-hole bubble.
Transport quantities can be expressed in terms of a generalized transport lifetime 
(see, e.g., Ref.~\onlinecite{pruschke:dmftNCA_HM95} or \onlinecite{merinoTransportDMFT00})
\begin{align}
  \label{eq:transportL}
  L_{\alpha\beta}=&\int\!d\o \left(-\frac{\partial f(\o)}{\partial \o} \right) \:\big[\tau_{xx}(\o)\big]^\alpha\:\o^{\beta-1}
\end{align}
where $f(\o)=1/(e^{\o/T}+1)$ is the Fermi function. The generalized transport lifetime is given by
\begin{align}
  \label{eq:tau}
  \tau_{xx}(\o)=&\frac{1}{N\pi^2}\sum_\k  \left(\frac{\partial t_\k}{\partial k_x} \right)^2 \left[\text{Im}G(t_\k,\o-i0^+)\right]^2
  \\
  =&\frac{1}{\pi^2}\int\!d\epsilon\: \tilde\rho_0(\epsilon)\left[\text{Im}G(\epsilon,\o-i0^+)\right]^2
\quad.
\end{align}
The function
\begin{align}
  \label{eq:rtil}
  \tilde\rho_0(\epsilon)&=\frac{1}{N}\sum_\k  \left(\frac{\partial t_\k}{\partial k_x} \right)^2\:\delta(\epsilon-t_\k)
\end{align}
can be calculated for the cubic lattices and turns out to be a smooth function
where the singularities of $\rho_0$ are removed due to the derivative.

Only in the Fermi liquid regime is $\tau_{xx}(\o)\sim 1/\text{Im}\Sigma^U(\o-i0^+)$
and the linearized  Boltzmann theory is recovered.\cite{AshcroftMermin:BOOK}

The resistivity
\begin{align}
  \label{eq:res}
  \varrho=\frac{\varrho_0}{L_{11}}
\end{align}
with $\varrho_0=\frac{2\hbar a}{\pi e^2}$ ($a$ is the lattice spacing, $e$ the electronic charge) 
is shown in 
Fig.~\ref{fig:hubb_2dscnnn_asymTrans}(b) for different filling. Typical characteristics
of strongly correlated systems can be observed for not too large doping.
The resistivity  $\varrho$ changes from insulating (or semi-conducting) behavior
at high temperatures to metallic behavior at low temperatures 
due to the emergence of (coherent) low-energy 
quasiparticles (see inset).\cite{merinoTransportDMFT00} 
The high-temperature resistivity is much larger than the value expected 
from the Ioffe-Regel condition valid for usual metals, where $\varrho$
is bound
by a minimal scattering length of the order of the lattice 
spacing.\cite{gunnarsonResitivity03}
At intermediate temperatures $0.1 \lesssim T\lesssim0.5$ linear regimes 
with increasing slope for  decreasing doping can be recognized,
in accord with earlier
studies.\cite{jarrell:HM93b,pruschke:dmftNCA_HM95}

The log-log plot  (main panel) reveals that the low 
temperature  Fermi liquid behavior
$\varrho(T)\sim T^2$ is found
for $n=1$ and above (not shown).\footnote{The reason, why we observe a quadratic temperature 
dependence, which is characteristic for Fermi liquids in
dimensions $d>2$ and not the two-dimensional form
$\varrho\sim T^2 \ln(1/T)^2$ is  not completely clear to us at present.
It is probably related to the neglect of vertex correction within DMFT. 
Ignoring the momentum dependence in the two-particle vertex leads to a 
violation of momentum conservation at internal vertices.\cite{jarrell:QMCMEMNonLocalDMFT01} 
The resultant mean-field decoupling in the Bethe-Salpeter 
equations\cite{jarrell:symmetricPAM95,schmitt:sus05,schmittPhD08} for the 
two-particle current-current correlation function introduces additional
averages over the Brillouin zone and mimics an effectively higher 
dimensionality for transport quantities.}
A $T^2$-dependency is also observed for $n=0.92$.
But for the intermediate values $n=0.98$ and $n=0.96$ a qualitative 
different behavior is found.  $\varrho(T)$ 
more closely resembles power-laws with exponents $1$ and $1.5$, respectively.
The van Hove singularity is close to the Fermi level and 
the enhanced scattering increases the resistivity.

In summary, the interplay between the reduction of phase space 
volume for scattering and the qualitative changes in the 
frequency dependence of the self-energy within the temperature window 
causes the qualitative changes in the transport properties.

%%%%%%%%%%%%%%%%%%%%%%%%%%%%%%%%%%%%%%%%%%%%%%%%%%%%%%%%%%%%%%%%%%
\section{Relevance for the theory of cuprate superconductors} 
\label{sec:susNFLRelPG}

The $t$-$t'$-Hubbard model might be suitable
to describe the low-energy physics of cuprate superconductors.
Even though DMFT 
is non-perturbative and includes nontrivial \textit{local} many-body  correlations, it 
represents a rather poor approximation for the two-dimensional cuprate layers
due to the neglect of \textit{nonlocal} correlations. 
Therefore, we want to point out that we neither try to model the
cuprate superconductors nor can our results be directly transferred 
to those systems. 

However, even 
in more appropriate theories like cluster-DMFT\cite{maier:QuantumCluster05}
the van Hove singularities associated with the lattice structure are retained
and therefore the findings of  this work 
have some bearings on the  pseudogap and strange metal
phase of cuprate superconductors.
In the following we will elaborate on some of these aspects.

The van Hove singularity provides an additional mechanism for
kinks\cite{lanzaraPhonoKink01,shenPhonHTC02,grafSCSepHTC07,vallaHEK07,macridinHighEnergyKinkHM07,raasCollectiveMode09,byczukKinks09}
in the real part of the self-energy [see Fig.~\ref{fig:hubb_2dsc_sym}(b)]. 
The location of the kinks  is temperature dependent which should 
make it possible to discriminate these type of kinks in experiment or 
calculations.

\begin{figure}[t]
  \includegraphics[width=60mm]{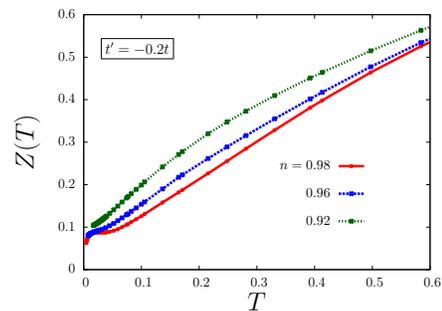}
  \caption{Generalized quasiparticle weight 
    for the Hubbard model on a square lattice with
    next-nearest neighbor hopping $t'=-0.2t$ and $U=4.25$ 
    as function of temperature for 
    three different filling obtained from DMFT(ENCA) calculations
    and  Eq.~(\ref{eq:qpzTf2}).
  } 
  \label{fig:qpz}
\end{figure}

The non-Fermi liquid  self-energy induces an unusual
temperature dependence of the generalized quasiparticle weight\cite{serene:QuasipartHM91} 
\begin{align}
\label{eq:qpzTf1}
  Z(T)&=\frac{1}{1-\frac{\text{Im}\Sigma^U(i\o_0)}{\o_0}}\\
  \label{eq:qpzTf2}
  &=\frac{1}{1+\int \frac{d\o}{\pi}\frac{\text{Im}\Sigma^U(\o-i0^+)}{\o^2+(\pi T)^2}}
  \quad,
\end{align}
where $\o_0=\pi T$ is the smallest fermionic
Matsubara frequency (The form (\ref{eq:qpzTf1}) is useful when the self-energy is obtained
only at the imaginary Matsubara frequencies like in quantum Monte-Carlo
calculations).
The generalized quasiparticle weight  as calculated with Eq.~(\ref{eq:qpzTf2}) 
is shown in Fig.~\ref{fig:qpz} 
for the Hubbard model
on the square lattice with next-nearest neighbor hopping  
as function of temperature and three different
filling. 
All curves show a change in curvature at the temperature where the 
anomalous double-well structure emerges.
This behavior closely resembles that found in Ref.~\onlinecite{vidhyadhiraja2DHMNFL09}
for a crossover
from a marginal Fermi liquid at high temperatures to
a Fermi liquid at low $T$.
Since the self-energies do not show any signs of  marginal Fermi liquid behavior
(see the previous sections) this reveals a difficulty when interpreting
imaginary Matsubara data at finite temperatures.  Due to the integral in 
Eq.~(\ref{eq:qpzTf2}) the temperature
dependence of $Z(T)$ is not determined by the asymptotic  low-frequency form 
of Im$\Sigma^U(\o-i0^+)$ but by its average spectral weight in the temperature window $|\o|\lesssim \pi T$.
Therefore, the discrimination between different characteristic low-energy forms of $\Sigma^U(\o-i0^+)$
given  the self-energy at imaginary Matsubara frequencies seems rather delicate
at finite temperature. 

The partial destruction of the Fermi surface (Fermi arcs) 
and the emergence of a pseudogap\cite{normanHTC98,kanigelHTC06,yangHTC08} 
is of current interest.
% are frequently studied within the two-dimensional Hubbard model
%and theories which  include nonlocal correlations are employed, 
%like cluster-extensions of  DMFT or the  dynamical vertex approximation.
These phenomena are usually explained to arises from the coupling
to strong nonlocal antiferromagnetic correlations.\cite{macridinPseudogapHM06,kyungPseudogap06,liebsch2DHM09,kataninDGA2dHM09}
In the present study no nonlocal fluctuations are included
but instead the van Hove singularity 
%Although t  presented calculations approach does not include such fluctuations, it 
produces similar features such as a
non-Fermi liquid maximum in Im$\Sigma^U(\o- i0^+)$
and the concomitant reduction of spectral weight [see Fig.~\ref{fig:hubb_gf_2dsc_asym}(a)],
However, a true pseudogap is not observed
in our approach indicating the importance of non-local 
fluctuations.

The existence of Fermi arcs and the asymmetry for 
momentum vectors along the nodal  $\gv{k}=\lambda (\pi,\pi)$ ($0\leq\lambda\leq 1$) and anti-nodal $\gv{k}=\lambda (0,\pi)$ 
direction is natural in the present scenario. The flat parts in the dispersion 
relation occur at the $X$ point in the Brillouin zone where 
the low-temperature quasiparticles are most strongly affected 
and consequently the Fermi liquid description breaks down there first.\cite{kataninQPRG04,roheNFL05}
Also the stability of the induced non-Fermi liquid behavior up to a doping
of the order of $20\%$ which was encountered in
other studies\cite{liebsch2DHM09} can be explained by the pinning of the
van Hove singularity to the Fermi level.

Unarguably, nonlocal correlations are vital for the understanding 
of cuprate superconductors. But as the above aspects suggest,
some of the non-Fermi liquid  signatures 
%observed in the calculations for the $2d$ square lattices are 
can be produced -- or at least aided -- by the presence
of a van Hove singularity in the vicinity of the Fermi surface.
This is especially important for lattices without nesting or with 
frustration, 
%%--- such as the square  lattice with next-nearest neighbor hopping $t'$ ---
where nonlocal antiferromagnetic spin-fluctuations 
are usually suppressed and thus might be too weak to produce  
non-Fermi liquid  signatures.
Therefore, we argue that the interpretation solely in terms of collective 
modes which couple to the electronic degrees of freedom is too simplified
and the influence of the saddle points in the Brillouin zone has to be 
reconsidered.

%%%%%%%%%%%%%%%%%%%%%%%%%%%%%%%%%%%%%%%%%%%%%%%%%%%%%%%%%%%%%%%%%%%%%%%%%%%%%
\section{Conclusions}

In the present work we have focused on 
the influence of a van Hove singularity in the non-interaction
density of states  on the low temperature properties of the Hubbard model. 
In order to include non-perturbative correlation effects we used the DMFT 
to calculate the spectral function and self-energy.   
As impurity solvers we employed the enhanced non-crossing approximation
for finite temperatures $T>0$ and the numerical renormalization
group for $T=0$.  Both methods yield dynamic quantities directly 
on the real frequency axis, which avoids the inaccuracies connected with
a numerical analytic continuation of imaginary time data.

The van Hove singularity causes profound changes of
the low-energy Fermi liquid properties usually encountered within DMFT for the 
Hubbard model. For the strongly correlated metal close to the Mott Hubbard 
metal-insulator transition the imaginary part of the self-energy
develops an unusual double-well structure at finite temperatures.
This anomalous  structure appears at temperatures on the order 
of the  many-body scale $T_0$  which also determines its energy spread. 
It originates from an enhanced scattering of 
the low-energy quasiparticles at the  saddle points in the dispersion relation 
associated with the van Hove singularity. 

Using a model DOS, we have shown analytically 
that the non-Fermi liquid signature
in the quasiparticle self-energy is directly linked 
to  non-analytic logarithmic contributions to 
the lattice scattering matrix.
As a consequence the medium for the effective impurity develops
a dip at the van Hove singularity. In case of a logarithmically diverging
non-interacting DOS this dip produces a soft-gap in the effective medium 
at $T=0$.

At zero temperature the interacting spectral function for a square lattice
exhibits a logarithmic divergence for half-filling. The  system is
well characterized by a generalized Fermi liquid where
the quasiparticle weight remains
finite and the imaginary part
of the self-energy  vanishes at
the Fermi level. But the frequency dependent scattering rate 
increases with a non-Fermi liquid exponent $\approx 1.5\neq 2$.
Upon doping a pinning of the van Hove singularity to the Fermi level
is observed. Its signature in the spectral function is always 
renormalized  toward the Fermi level and stays close to it
up to rather large doping $\delta\lesssim 0.2$.

For the square lattice with a finite next-nearest neighbor hopping 
$t'=-0.2t$ the van Hove singularity in the non-interacting
DOS is located below the Fermi level and moved toward it  
upon increasing hole-doping. This trend is also 
observed in the anomalous maximum in the self-energy.

In case of half-filling and electron doping, as well as for large 
hole doping, the zero-frequency quasiparticle scattering rate exhibits 
a low temperature dependence which is consistent with  that of a 
two-dimensional Fermi liquid. In between these fillings the decrease 
is much slower and follows a power law with exponent close to one. 

A similar signature of the van Hove singularity was found in the
resistivity. A Fermi liquid  $T^2$  dependence emerged for 
$n\gtrsim 1$ and $n\lesssim 0.92$, while for $0.92\lesssim n\lesssim 1$
the resistivity was enhanced and a decrease with exponents less than two 
was observed.

The findings of this work bear some implications on theories
for cuprate 
superconductors.
Even thought no nonlocal fluctuations were included, we still produced 
qualitative features usually only obtained within more advanced 
theories where those correlations are incorporated. This raises the questions concerning 
the origin and physical mechanism behind such features and  
we think the role of the van Hove singularity in connection with 
strong correlations  should be further explored in the future.

\begin{acknowledgments}
  The author acknowledges very fruitful discussions with 
  N.\ Grewe, F.\ B.\ Anders, T.\ Jabben,
  E.\ Jakobi and F.\ G\"uttge.
  The author especially thanks F.\ B.\ Anders for providing him with
  his NRG code with which the zero temperature calculations of this work
  were done.
  This work was supported by the Deutsche
  Forschungsgemeinschaft under Grant No. AN 275/6-2.
\end{acknowledgments}

\appendix*

\section{Details of the model DOS calculation}
\label{appCusp}

The scattering matrix can be calculated analytically via Eq.~(\ref{eq:scatter}) for 
any $\tilde G$ (omitting the arguments),
\begin{align}
  \label{eq:TToyDos}
  T&  = \int_{-W}^W\!dx\: \frac{ x^2\:\rho^{\text{cusp}}_{\alpha}(x)}{\tilde{G}^{-1}-x}
  \\  \label{eq:TToyDos1}
  & \overset{\alpha>0}{=} \frac{1+\alpha}{\alpha}
  \frac{1}{\tilde{G}}\left[
    -1+\frac{1}{\tilde{G} W}\mathrm{atanh}(\tilde{G} W) 
  \right. 
  % \qquad , \quad \alpha>0
  \\\notag&
  \left.   -\frac{(\tilde{G} W)^2}{3+\alpha} \:
    \: {}_2 \text{F}_1\left(1,\frac{3+\alpha}{2};\frac{5+\alpha}{2};(\tilde{G}W)^2\right)
  \right]
  \\
  &\overset{\alpha=0}{=} \frac{1}{\tilde{G}}\left[
    -1 +\frac{\text{Li}_2(\tilde{G}W) -\text{Li}_2(-\tilde{G}W)}{2\tilde{G}W}
  \right]
% \qquad , \quad \alpha=0
  \quad,
\end{align}
where ${}_2$F$_1$ is the Gauss's hypergeometric function and Li$_2$ the Dilogarithm.
Equation~(\ref{eq:mediumSC}) then yields  the effective
medium
\begin{widetext}
\begin{align}
  \Gamma&
  \label{eq:hubToyLinDosAlpha}
  \overset{\alpha>0}{=}
  \frac{1}{\tilde{G}}-
  \frac{W \alpha }{
    (1+\alpha)\mathrm{atanh}(\tilde{G} W)-\tilde{G}W
    -\frac{1+\alpha}{3+\alpha}\big(\tilde{G}W\big)^3\;{}_2\text{F}_1\Big(1,\frac{3+\alpha}{2};\frac{5+\alpha}{2};(\tilde{G}W)^2\Big)
  }
  \\  \label{eq:hubToyLinDos0A}
  &\overset{\alpha=0}{=}
  \frac{1}{\tilde G} + \frac{2 W}{\text{Li}_2(-\tilde{G}W) - \text{Li}_2(\tilde{G}W)}
  \quad.
\end{align}
\end{widetext}

For $\alpha=0$, Eq.~\eqref{eq:hubToyLinDos0} already constitutes the
final expression, while for the $\alpha=1,2$  equation~\eqref{eq:hubToyLinDosAlpha} 
can  be further  simplified.   
  %%%%%%%%%%%%%%%%%%%%%%%%%%%%%%%%%%%%%% 

For $\alpha=1$ the hypergeometric function reduces to
  \begin{align}
    \label{eq:hubHyperGeo1}
    {}_2\text{F}_1\left(1,2;3;z^2\right)&=-\frac{2}{z^4}[z^2+\ln(1-z^2)]
    \quad,
  \end{align}
  leading to the effective medium displayed in Eq.~(\ref{eq:hubToyLinDos1}).
  In case of $\alpha=2$
  \begin{align}
    \label{eq:hubHyperGeo2}
    {}_2\text{F}_1\left(1,\frac{5}{2};\frac{7}{2};z^2\right)&=
    -\frac{5}{3}\frac{1}{z^5}[3 z+z^3-3 \mathrm{atanh}(z)]
    \quad,
  \end{align}
  yields the effective medium of Eq.~(\ref{eq:hubToyLinDos2}).

Assuming a Fermi liquid as in Eq.~(\ref{eq:flSig}) and taking the limit 
$\gamma\to0$ the effective media take the approximate low-energy forms, $\o\ll W=1$
($Z=1$, $T=0$, $\tilde\e=0$)
\begin{align}
  \label{eq:tta0w0}
  \text{Im}\Gamma_{\alpha=0}(\omega-i0^+)&\approx -\frac{2}{\pi}\frac{\ln|\omega|}{\frac{\pi^2}{4}+(\ln|\omega|)^2}
  \\
  \label{eq:tta1w0}
  \text{Im}\Gamma_{\alpha=1}(\omega-i0^+)&\approx \frac{\pi(1-|\o|)}{\pi^2(1-|\o|)^2+4\omega^2(1-\ln|\omega|)^2}
  \\
  \label{eq:tta2w0}
  \text{Im}\Gamma_{\alpha=2}(\omega-i0^+)&\approx
  \frac{\pi}{3}\frac{1-\o^2}{\frac{\pi^2}{4}+(4-\frac{\pi^2}{2}) \o^2 }
  \quad.
\end{align}
For $\alpha=0$ the effective medium reaches zero at zero energy. For $\alpha=1$ it 
saturates at a value $1/\pi$ but the non-analytic cusp still remains, while for $\alpha=2$ 
the cusp is removed and Im$\Gamma_{\alpha=2}(-i0^+)=4/(3\pi)$.

%\bibliography{masterbib}

%%%%
\end{document}